\documentclass[12pt,letterpaper]{article}

\usepackage{graphicx,array,soul}
\usepackage{url}
\usepackage{color}
\usepackage{latexsym}
\usepackage{amsthm}
\usepackage{amsmath}
\usepackage{amssymb}
\usepackage{amsfonts}
\usepackage[numbers,sort&compress]{natbib}
\usepackage{bm}
\usepackage{slashed}
\usepackage{mathrsfs}
\usepackage{enumerate}
\usepackage{tikz}
\usepackage{siunitx}
\usepackage{mdframed}
\usepackage{setspace}  
\usepackage{esvect}
\usepackage{multirow}
\usepackage[hidelinks]{hyperref}

\def\Mpl{M_{\rm pl}}
\def\tauosc{\tau_{\rm osc}}
\def\lsim{\mathrel{\rlap{\lower4pt\hbox{\hskip1pt$\sim$}}
     \raise1pt\hbox{$<$}}}
\def\gsim{\mathrel{\rlap{\lower4pt\hbox{\hskip1pt$\sim$}}
     \raise1pt\hbox{$>$}}}

\newcommand{\mc}{\mathcal}
\newcommand{\bv}[1]{{\bf #1}}

\usepackage{natbib}
\begin{document}

\title{ 
{\bf Challenges in Interpreting the NANOGrav 15-Year Data Set as Early Universe Gravitational Waves Produced by ALP Induced Instability}
\author{\large Michael Geller$^{\,a}$, Subhajit Ghosh$^{\,b}$, Sida Lu$^{\,c}$, and Yuhsin Tsai$^{\,b}$}
\date{\small \it 
$^a$School of Physics and Astronomy, Tel Aviv University, Tel Aviv 69978, Israel \\
$^b$Department of Physics and Astronomy, University of Notre Dame, South Bend, IN 46556, USA \\
$^c$Institute for Advanced Study, The Hong Kong University of Science and Technology, \\Clear Water Bay, Kowloon, Hong Kong S.A.R., P. R. China \\
}
}

\maketitle

\setlength{\parskip}{0.2ex}

\begin{abstract}
In this paper, we study a possible early universe source for the recent observation of a stochastic gravitational wave background at the NANOGrav pulsar timing array. The source is a tachyonic instability in a dark gauge field induced by an axion-like particle (ALP), a known source for gravitational waves. 
We find that relative to the previous analysis with the NANOGrav 12.5-year data set, the current 15-year data set favors parameter space with a relatively larger axion mass and decay constant. This favored parameter space is heavily constrained by $\Delta N_{\rm eff}$ and overproduction of ALP dark matter. While there are potential mechanisms for avoiding the second problem, evading the $\Delta N_{\rm eff}$ constraint remains highly challenging. In particular, we find that the gravitational wave magnitude is significantly suppressed with respect to the gauge boson dark radiation, which implies that successfully explaining the NANOGrav observation requires a large additional dark radiation, violating the cosmological constraints. Satisfying the $\Delta N_{\rm eff}$ constraint will limit the potential contribution from this mechanism to the observed signal to at most a percent level. 
\end{abstract}

\section{Introduction}\label{sec:introduction}

The recently released common gravitational wave spectrum measurements by NANOGrav~\cite{NANOGrav:2023gor}, CPTA~\cite{Xu:2023wog}, EPTA~\cite{Antoniadis:2023aac} and PPTA~\cite{Reardon:2023gzh} report a monumental 3$\sigma$-level observations of the Hellings-Downs curve. This is a strong hint of an isotropic cosmic stochastic gravitational wave background (SGWB) which opens a new window into the universe.
Identifying the source of these gravitational wave signals is the first step to look into this new window, and it is particularly crucial to understand if any beyond the Standard Model physics are involved.
In addition to the conventional sources of binary supermassive black hole mergers~\cite{Phinney:2001di,Sesana:2012ak,Chen:2018rzo,Zhu:2011bd}, the possible new physics relevant for the nanohertz SGWB include cosmic phase transitions~\cite{Freese:2022qrl,Levi:2022bzt,Ellis:2020nnr,Lewicki:2020azd,Lewicki:2020jiv,Hindmarsh:2019phv,Caprini:2019egz,Zu:2023olm,Franciolini:2023wjm,Megias:2023kiy,Han:2023olf,Fujikura:2023lkn,Li:2023bxy,Ashoorioon:2022raz,Athron:2023mer,Ghosh:2023aum,Samanta:2020cdk,DiBari:2023upq}, cosmic strings~\cite{Auclair:2019wcv,Blasi:2020mfx,Ellis:2020ena,Ferrer:2023uwz,Wang:2023len,Ellis:2023tsl}, domain walls~\cite{Saikawa:2017hiv,Ferreira:2022zzo,Kitajima:2023cek,Bai:2023cqj,Lazarides:2023ksx,Barman:2023fad,Sakharov:2021dim,King:2023cgv,Ge:2023rce}, secondary GWs from primordial density perturbations~\cite{Pi:2020otn,Baumann:2007zm,Dandoy:2023jot,Choudhury:2023rks,Cai:2023dls}, solitons~\cite{Lozanov:2023aez,Lozanov:2022yoy,Broadhurst:2023tus}, ultralight scalars~\cite{Yang:2023aak,Guo:2023hyp,Anchordoqui:2023tln}, dark matter spikes around supermassive black holes~\cite{Shen:2023pan}, primordial magnetic field~\cite{Li:2023yaj}, cosmic neutrinos~\cite{Lambiase:2023pxd} and non-Gaussianity~\cite{Franciolini:2023pbf,Liu:2023ymk}. In this work, we examine the possibility that this SGWB is sourced by the dynamics of axion-like particles (ALPs).

Axions are pseudo-Goldstone bosons first proposed as the solution to the strong CP problem~\cite{Peccei:1977hh,Peccei:1977ur}, and their quanta were soon realized to also serve as viable dark matter (DM) candidates~\cite{Abbott:1982af,Preskill:1982cy,Dine:1982ah,Co:2017mop}.
ALPs, as suggested by their name, inherit and generalize the cosmological phenomenology of axions, while do not necessarily have a connection to the strong CP due to its shift symmetry-breaking mass.
From now on, we will use axions and ALPs interchangeably, with no connection to the strong CP problem. 
ALPs can be responsible for the cosmic inflation and  reheating~\cite{Freese:1990rb,Dimopoulos:2005ac,Anber:2009ua}, address the hierarchy problem~\cite{Hook:2016mqo,Fonseca:2019ypl,Graham:2015cka} and also generically arise in various string theory models~\cite{Chadha-Day:2021uyt,Cicoli:2012sz}.
Specifically, when coupled to massless or very light gauge modes like the dark photons, the ALP field's rolling can exponentially amplify a specific helicity of these gauge modes, a process known as ``tachyonic instability''. This can serve as an efficient mechanism for particle production in the universe. 
This mechanism has been studied in the context of reheating~\cite{Anber:2009ua,Anber:2012du}, production of dark photon DM~\cite{Co:2018lka,Agrawal:2018vin,Dror:2018pdh} and depletion of axion DM to avoid over-closure~\cite{Agrawal:2017eqm}.
Recent studies show that the drastic enhancement of the vector fields through this mechanism can conveniently serve as a source of GWs that may be visible on interferometers or pulsar timing arrays (PTAs)~\cite{Machado:2018nqk,Machado:2019xuc}, and generate CMB B-modes if the ALP is light enough~\cite{Geller:2021obo}.

In this work, we analyze the interpretation of the recent GW observation by the NANOGrav collaboration in their 15-year data set~\cite{NANOGrav:2023gor} (hereafter NG15) as primordial GWs induced by tachyonic instability at around MeV temperatures\footnote{A similar analysis was carried out in Ref.~\cite{Figueroa:2023zhu} without addressing the constraints we derive here. Additionally, a similar production mechanism during the inflationary epoch for the NG15 SGWB was considered in Ref.~\cite{Niu:2023bsr,Murai:2023gkv}.}. This interpretation was considered by Ref.~\cite{Madge:2023cak} for the earlier NANOGrav 12.5-year data set (NG12)~\cite{NANOGrav:2020bcs} and the IPTA's second data release~\cite{Antoniadis:2022pcn}. The authors find that the combined data favors a region that is in mild tension with $\Delta N_{\rm eff}$ constraints, i.e. the constraints on the effective light degrees of freedom at the big-bang nucleosynthesis (BBN)~\cite{Adshead:2020htj} and cosmic microwave background (CMB)~\cite{Planck:2018vyg}. In this paper, we will show that with the updated NG15 data, this interpretation is in severe tension with $\Delta N_{\rm eff}$ constraints, which is virtually unavoidable as we explain below. This is explained by the NG15 data preferring a higher peak magnitude, especially with the improvement of the data at higher frequency bins.

In Ref.~\cite{Madge:2023cak}, the analysis is carried out by using a formula fitted to their lattice result (see also Ref.~\cite{Ratzinger:2020oct}). In this analysis, we use both the linear calculation and the fitted lattice spectrum, which differ significantly in the predicted frequency of the gravitational wave, but not in its magnitude. Therefore, while the parameter regions favored by the NG15 data in our two approaches are different, the expected $\Delta N_{\rm eff}$ values for both parameter regions are very similar. The exclusion we derive here is therefore robust.

In the following, we explain the two major constraints on this scenario. These are the over-closure constraints from the massive ALP field, and the $\Delta N_{\rm eff}$ constraints from the massless gauge modes. 
These two constraints arise from the same origin: to produce the observed GW signal, the initial energy density of the ALP field must be large enough. 
While a substantial portion of this energy density is eventually converted into dark gauge field quanta, the remaining energy density of the ALP field undergoes matter-like redshift. However, across the entire parameter space, this energy density significantly exceeds the observed DM relic abundance, resulting in an over-closure constraint. It is conceivable that this constraint could be circumvented by transferring the matter-like energy density into radiation or kinetic energy, which undergoes faster redshift, through additional model building. The $\Delta N_{\rm eff}$ constraint comes from the energy budget in the dark gauge field quanta that behave as dark radiation. As we show in this work, the energy density of gauge quanta required to produce the GW signal far exceeds the constraint on $\Delta N_{\rm eff}$. Since both the GW and the dark radiation densities redshift in the same way, it is much harder to avoid this constraint.
Consequently, the aforementioned explanation of the NG15 signal is disfavored, and potential GW signals from the ALP-induced tachyonic instability are expected to be a few orders of magnitude below the observed signal.

Despite no significant evidence being found by the analysis of the NANOGrav Collaboration at the current experimental sensitivity~\cite{NANOGrav:2023tcn}, an anisotropic GW background remains a very interesting possibility and could naturally arise from, {\it e.g.} the fluctuations of the GW sources like supermassive black hole mergers~\cite{Mingarelli:2017fbe}. Here we ignore this possibility and base our analysis on the assumption of an isotropic GW signal.

The organization of the paper is as follows: In Sec.~\ref{sec:model} we first briefly review the ALP model, the tachyonic production of dark radiation from the dynamic of ALPs, and how GWs are sourced in this process. Then in Sec.~\ref{sec:results} we describe the setup of our numeric calculation and show the model parameter space where the produced GW reasonably explains the NG15 results. In this context, we discuss the severity of the constraints from the cosmic dark matter relic abundance and $\Delta N_{\rm eff}$. After this, we discuss some future perspectives on this model and conclude.

\section{The Instability-induced Gravitational Waves}\label{sec:model}

\subsection{Tachyonic production of dark photons}\label{sec:tachyonic}

Consider the Lagrangian of an axion coupling to the gauge boson of a $U(1)$ gauge symmetry
\begin{align}
\mc{L}=\dfrac{1}{2}\partial_\mu\phi \partial^\mu\phi-V(\phi)-\dfrac{1}{4}X_{\mu\nu}X^{\mu\nu}-\dfrac{\alpha}{4f}\phi X_{\mu\nu}\tilde{X}^{\mu\nu}\,,
\end{align}
where $\phi$ is the axion field, $f$ is the axion constant, $X^{\mu\nu}=\partial^\mu X^\nu - \partial^\nu X^\mu$ is the exterior derivative of the massless dark photon field $X^\mu$ and $\alpha$ is the axion-dark photon coupling.
The axion potential is chosen to have the generic quadratic form of $V(\phi)=\frac{1}{2}m^2\phi^2$ where $m$ is the axion mass, which can be seen as the leading order expansion of the axion-like potential $\Lambda^4\cos(\phi/f)$.
At the leading order, we ignore the spatial inhomogeneity of the axion field so that the axion field $\phi$ does not carry a momentum dependence. We discuss the full solution including the ALP inhomogeneities in Sec.~\ref{sec:nonlin}.
The equation of motion of $\phi$ can then be written as
\begin{align}\label{eq:axion_eom}
\phi^{\prime\prime}+2aH\phi^\prime+a^2\frac{\partial V}{\partial \phi}=\frac{\alpha}{f}a^2\bv{E}\cdot\bv{B}\,,
\end{align}
where $a$ is the scale factor of the FRW metric, and $H$ is the Hubble parameter.
The prime $^\prime$ indicates a derivative with respect to the conformal time $\tau$.
The $\bv{E}$ and $\bv{B}$ on the right-hand side are the corresponding electric and magnetic fields of the dark photons.
As in the case of free axions, $\phi$ receives a Hubble damping when the Hubble parameter is large, 
 and the field only starts rolling when $H\sim m$.
As a convention, we define the oscillation time $\tau_{\rm osc}$ to be the conformal time where $H(\tau_{\rm osc})=m$, and $a_{\rm osc}$ to be the scale factor at that time.
At the beginning of the rolling, the $\bv{E}\cdot\bv{B}$ term is negligible, and the axion dynamics is the same as that of free axions.
After the initiation of tachyonic production of the dark photon, as explained in detail below, the $\bv{E}\cdot\bv{B}$ term will introduce increased friction to the axion motion, leading to energy transfer from axions to dark photons. 
The dark photon field is quantized as
\begin{equation}\label{eq:dark_photon}
\begin{aligned}
&X_i(\bv{x}) = \int\mathcal{D}k\left(\epsilon_{+i}(\bv{k})v_+(\tau,k)\hat{{\bf a}}_+(\bv{k})e^{i\bv{k}\cdot\bv{x}}+h.c.\right),\\
&X_0(\bv{x}) = 0\,,
\end{aligned}
\end{equation}
where we take the Coulomb gauge and write $\mathcal{D}k\equiv d^3k/(2\pi)^3$. The polarization vector $\epsilon_\pm(\bv{k})$ satisfy $\bv{k}\cdot\bv{\epsilon}_\pm=0$, $\bv{k}\times\bv{\epsilon}_\pm=\mp i k\bv{\epsilon}_\pm$, $\bv{\epsilon}_\pm\cdot\bv{\epsilon}_\pm=0$, $\bv{\epsilon}_\pm\cdot \bv{\epsilon}_\mp=1$~\cite{Anber:2012du}.
Due to the spatial isotropy, the mode functions $v_\pm$ should depend only on the magnitude of the wave number $k=\lvert\bv{k}\rvert$, and they have the equation of motion
\begin{align}\label{eq_v}
v^{\prime\prime}_\pm(k,\tau)+\omega^2_\pm(k,\tau) v_\pm(k,\tau)=0,
\end{align}
with the dispersion relation $\omega^2_\pm(k,\tau)=k^2\mp k\alpha\phi^\prime/f$. 

In Eq.~\eqref{eq_v}, it is evident that when the axion is rolling, one helicity of the gauge modes with momentum $0 < k < \alpha\lvert\phi^\prime\rvert/f$ will become tachyonic, i.e. $\omega^2<0$, resulting in an exponential amplification. The amplification of the $v_{+(-)}$ modes occurs when $\phi^\prime$ is positive (negative), and the enhancements on the two helicities alternate as the axion field rolls around the minima of $V(\phi)$. In this context, it is reasonable to expect that the helicity experiencing initial enhancement during the ALP field's rolling motion would be more amplified compared to the other, as it spends a longer duration within the tachyonic band.

However, the tachyonic enhancement of the dark photons is not perpetual. As the axion field undergoes redshift and spends less time rolling on either side of the potential, the available time for the gauge modes to amplify decreases, even if their momenta fall within the tachyonic band. The growth time scale of a gauge mode in the tachyonic band is about $\lvert\omega_\pm\rvert^{-1}$, while the axion oscillation time scale is about $1/(am)$ as estimated from free rolling.
For a gauge mode to be enhanced we should require the former time scale to be shorter, such that there will be enough growth time for the gauge mode.
This suggests that the tachyonic production process should cease at the time $\tau_s$ where $a(\tau_s)/a_{\rm osc}\sim (\alpha/2)^{2/3}$~\cite{Machado:2018nqk}.

\begin{figure*}[t!]
\centering
\includegraphics[width=0.49\linewidth]{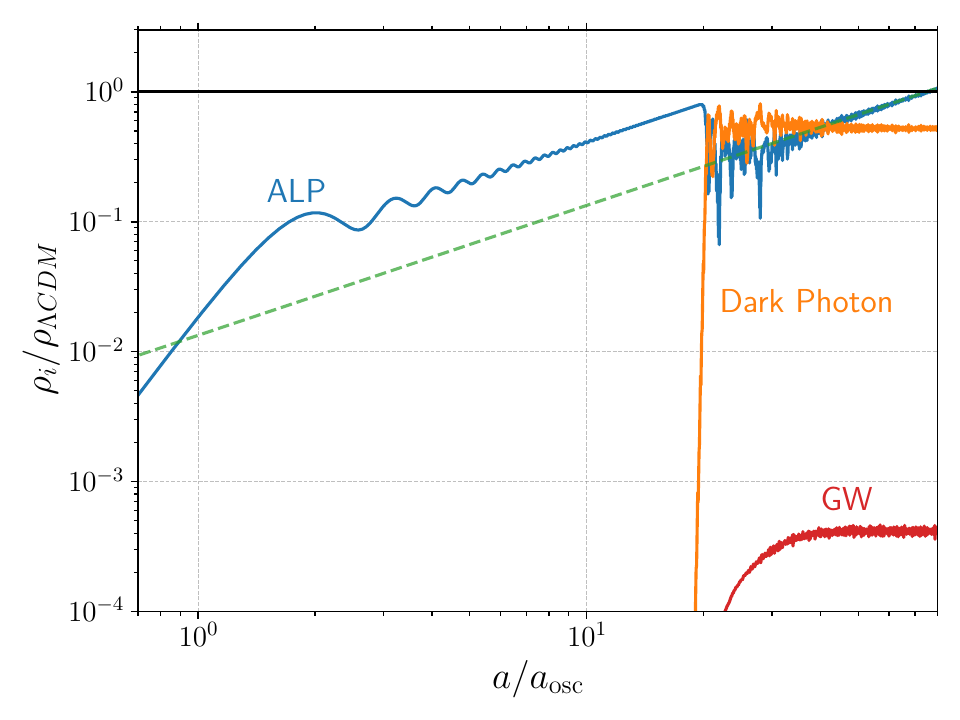}
\includegraphics[width=0.49\linewidth]{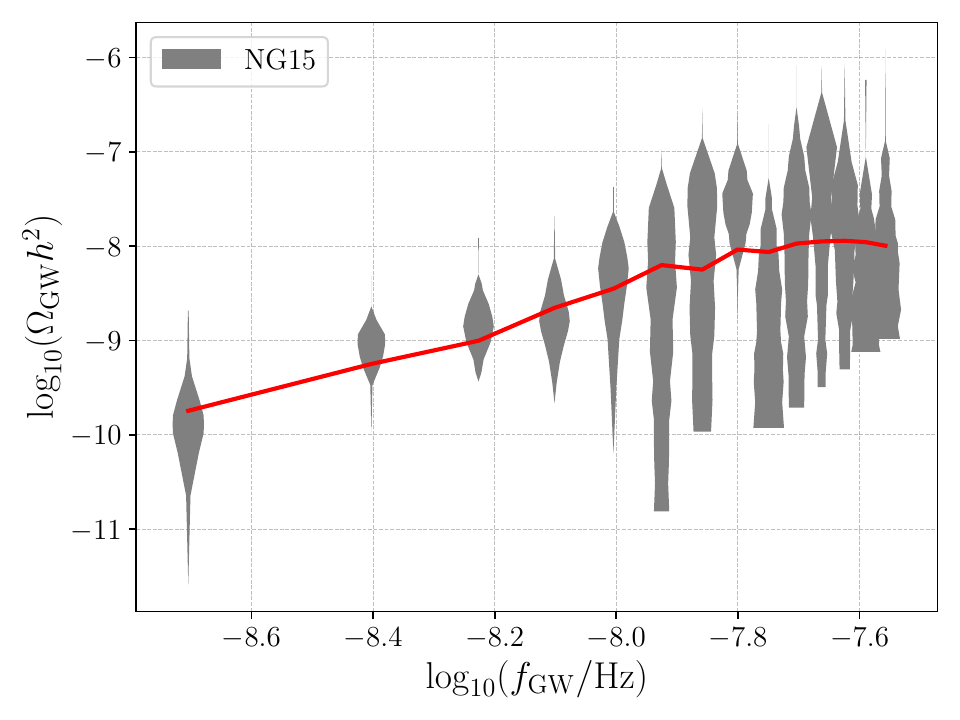} 
\caption{
An illustration of the tachyonic production mechanism and the induced GWs from the linear-order calculation, where the model parameters are set to $m\approx 8\times 10^{-12}$eV, $f\approx 8\times 10^{17}$ GeV, $\alpha=50$. 
{\it Left:} The evolution of the ALP (blue), dark photon (orange), and GW (red) energy density, normalized by the total energy density of the standard $\Lambda$CDM cosmology. 
The dashed green line is proportional to $a$, showing that ALPs now evolve like matter.
Soon after the ALP starts rolling, at $a/a_{\rm osc}\approx 20$ the tachyonic production of gauge modes is maximized and the ALP quickly dumps its energy into dark radiation.
After the tachyonic production, the ALP and dark photon continue to behave like ordinary matter and radiation.
On the other hand, this set of parameters has the ALP relic abundance overclose the universe and has too many radiation degrees of freedom at the BBN and CMB.
{\it Right:} The GW generated from the tachyonic instability (red curve), compared with the first 14 frequency bins of NG15's free spectrum density (gray shades), taken from Ref.~\cite{NANOGrav:2023gor}.
}
\label{fig:BM}
\end{figure*}

The left panel of Fig.~\ref{fig:BM} illustrates the evolution of this coupled ALP-dark photon system, where the energy density evolution of the ALPs ($\rho_\phi$), dark photons ($\rho_X$) and GWs ($\rho_{\rm GW}$) are shown.
The energy densities are normalized against the cosmic energy density of the standard $\Lambda$CDM cosmology. The ALP field starts rolling at $a_{\rm osc}$ and behaves as a freely rolling field before efficient radiation production occurs. 
It takes a few oscillations for the axion field to release a significant amount of its energy into GW and dark photon radiation (DR), depending on the size of $\alpha$. After the tachyonic production ceases, the remaining ALPs and dark photons start to redshift like ordinary matter and radiation. When solving the equations of motion for the ALP and dark photons, we adopt the assumption of a standard $\Lambda$CDM evolution for the Hubble parameter $H$ and scale factor $a$.
For the majority of the parameter space favored by the NG15 measured spectrum, this assumption holds reasonably when the tachyonic production is active. When $f$ and $m$ are large and $\alpha$ is small, on the other hand, there may be a mild violation where $\rho_\phi/\rho_{\rm \Lambda CDM}\gtrsim \mathcal{O}(1)$ during or at the end of the production. While note that this is also the parameter space suffering more strongly from the over-closure issue (see Fig.~\ref{fig:overclosure} in later texts), we do not expect our final conclusion to change significantly.

\subsection{The instability-induced GW spectra}
As the gauge modes undergo tachyonic amplification, they begin to act as sources for the SGWB. The generated metric tensor perturbation is given by
\begin{align}
h_{ij}(\bv{k},\tau)&=\dfrac{2}{\Mpl^2 a(\tau)}\int^\tau_{\tauosc} d\tau^\prime a(\tau^\prime)G(k,\tau,\tau^\prime)\Pi_{ij}(\bv{k}, \tau^\prime)\,,
\end{align}
where $\Pi_{ij}$ is the transverse traceless energy-momentum tensor of the gauge modes, and $G(k,\tau,\tau^\prime)$ is the Green's function of the tensor perturbation's equation of motion.
As our signal is generated during radiation domination, we explicitly have $G(k,\tau,\tau^\prime) = \sin(k(\tau-\tau^\prime))/k$.
The GW spectra are then calculated as
\begin{align}\label{eq:GW_spectra_1}
\dfrac{d\Omega_{\rm GW}(\tau)}{d\log k}=&\dfrac{d\rho_{\rm GW}(\tau)}{\rho_c d\log k}=\dfrac{1}{\rho_c}\dfrac{\Mpl^2 k^3}{8\pi^2 a^2(\tau)}\dfrac{\langle h^\prime(\bv{k},\tau)h^\prime(\bv{k^\prime},\tau)\rangle}{(2\pi)^3\delta(\bv{k}-\bv{k}^\prime)}\nonumber\\
=&\dfrac{1}{\rho_c}\dfrac{k^3}{2\pi^2\Mpl^2 a^4}\int^\tau_{\tauosc} d\tau^\prime_1 d\tau^\prime_2\, a(\tau^\prime_1)a(\tau^\prime_2)\mc{G}^\prime(k,\tau,\tau^\prime_1)\mc{G}^\prime(k,\tau,\tau^\prime_2)\Pi^2(\bv{k},\tau^\prime_1,\tau^\prime_2)\,,
\end{align}
where $\mc{G}^\prime(k,\tau,\tau^\prime)=\dfrac{d}{d\tau}G(k,\tau,\tau^\prime)-\dfrac{a^\prime(\tau)}{a(\tau)}G(k,\tau,\tau^\prime)$, and $\Pi^2(\bv{k},\tau^\prime_1,\tau^\prime_2)=\langle\Pi_{ij}(\bv{k}, \tau^\prime_1)\Pi_{ij}(\bv{k}, \tau^\prime_2)\rangle$ is the unequal time correlator of the gauge modes' energy-momentum tensor. 
It can be shown that~\cite{Machado:2018nqk}
\begin{align}
\Pi^2({\bf k},\tau,\tau^\prime)&=2\sum_{\lambda_1,\lambda_2=\pm}\int\dfrac{d^3q}{(2\pi)^3}\lvert\Theta_{\lambda_1\lambda_2}({\bf k}-{\bf q},{\bf k})\rvert^2\mathcal{S}_{\lambda_1\lambda_2}({\bf q},{\bf k},\tau)\mathcal{S}^\ast_{\lambda_1\lambda_2}({\bf q},{\bf k},\tau^\prime)\,,\\
\mathcal{S}_{\lambda_1\lambda_2}({\bf q},{\bf k},\tau)&=-\dfrac{1}{a^2(\tau)}\left[\lambda_1\lambda_2\lvert {\bf q}\rvert \lvert {\bf k}-{\bf q}\rvert v_{\lambda_1}({\bf q},\tau)v_{\lambda_2}({\bf k}-{\bf q},\tau)+v^\prime_{\lambda_1}({\bf q},\tau)v^\prime_{\lambda_2}({\bf k}-{\bf q},\tau) \right]\,,
\end{align}
where
\begin{align}
\lvert\Theta_{\lambda_1\lambda_2}({\bf k}-{\bf q},{\bf k})\rvert^2&=\lvert\Theta^+_{\lambda_1\lambda_2}({\bf k}-{\bf q},{\bf k})\rvert^2+\lvert\Theta^-_{\lambda_1\lambda_2}({\bf k}-{\bf q},{\bf k})\rvert^2\,,\\
\lvert\Theta^\lambda_{\lambda_1\lambda_2}({\bf k}-{\bf q},{\bf k})\rvert^2&=\dfrac{1}{16}\left(1+\lambda\lambda_1\dfrac{\bv{k}\cdot\bv{q}}{\lvert\bv{k}\rvert\lvert\bv{q}\rvert}\right)^2\left(1+\lambda\lambda_2\dfrac{\bv{k}\cdot(\bv{k}-\bv{q})}{\lvert\bv{k}\rvert\lvert\bv{q}\rvert}\right)^2\,,
\end{align}
is the projector function, and $v_\lambda$ are the quantized mode function as defined in~\eqref{eq:dark_photon}.
We are thus well-equipped to calculate the GW spectra.

The observed signals on PTAs correspond to the SGWB at the present time. Therefore, in Eq.~\eqref{eq:GW_spectra_1}, we should in principle integrate until the current age of the Universe, $\tau=\tau_0$. On the other hand, as discussed in the previous subsection, the tachyonic production of the gauge modes ceases at around $\tau\sim\tau_s$, and the active sourcing of GW stops thereafter. Hence, when evaluating Eq.~\eqref{eq:GW_spectra_1}, we only need to integrate up to a suitable point $\tau_{\rm max}$ after $\tau_s$. In other words, we approximate $\Pi^2(\tau^\prime_1,\tau^\prime_2)\to\Pi^2(\tau^\prime_1,\tau^\prime_2)\theta(\tau_{\rm max}-\tau^\prime_1)\theta(\tau_{\rm max}-\tau^\prime_2)$, where $\theta(z)$ represents the step function. After $\tau=\tau_{\rm max}$ the GW simply behaves like normal relativistic degrees of freedom and redshift as $a^{-4}$, {\it i.e.}, $\rho_{\rm GW}(k,\tau)=(a(\tau_{\rm max})/a(\tau))^4\rho_{\rm GW}(k,\tau_{\rm max})$, as indicated by the pre-factor in Eq.~\eqref{eq:GW_spectra_1}. The GW frequency $f_{\rm GW}$ redshifts as $a^{-1}$ and is related to the co-moving wave number $k$ as $k=2\pi a f_{\rm GW}$.

In the right panel of Fig.~\ref{fig:BM} we present the generated GW, and compare it with the first 14 frequency bins in the NG15  common-spectrum spatially-uncorrelated red noise-free spectrum, following the analysis in Ref.~\cite{NANOGrav:2023gor}.
If we only consider the GW spectrum and ignore other constraints, then the instability-induced GW is well capable of explaining the NG15 result. The mass $m$ and the coupling parameter $\alpha$ determine the peak momentum mode of the tachyonic band which translates to the peak frequency of the GW power spectrum. Ref.~\cite{Machado:2019xuc} suggests the peak frequency to be  given by
\begin{equation}\label{eq:GW_peak}
    f^0_{\rm GW, peak} \approx 2.0\times 10^{-8} \ {\rm Hz} \ \left(\dfrac{\alpha}{50}\right)^{\frac{2}{3}}\left( \frac{m} {4 \times 10^{-12} \ {\rm eV}} \right)^{\frac{1}{2}}\,.
\end{equation}
Using the above formula and the peak of the observed NG15 spectrum one can infer the relevant axion mass for instability-induced GWs to be a SGWB explanation. For $40<\alpha<100$ of our interest, we expect the ALP mass to lie approximately around $ m \approx [1.5 - 4.0] \times 10^{-12} \ {\rm eV}$. 

The peak amplitude of the GW spectrum on the other hand depends on the decay constant $f$ but not on the mass. At the onset of ALP rolling the total energy in the sector is given by $\rho_\phi = m^2 f^2/2$, and the cosmic energy density is $\rho_{\rm tot}=3H^2 \Mpl^2\propto m^2$ as $H\sim m$, and hence $\rho_\phi/\rho_{\rm tot}\propto f^2$ is independent of $m$. During tachyonic production, the ALP energy is dumped into dark photons, and thus we may reasonably expect $\rho_X / \rho_{\rm tot} \propto f^2$. This implies that the peak amplitude of $\Omega_{\rm GW} $, which is the two-point correlator of the tensor metric perturbation, scales as $f^4$ and is mostly independent of $m$~\cite{Machado:2018nqk,Geller:2021obo}, as it is sourced by $\rho^2_X$ (as shown in Eq.~\eqref{eq:GW_spectra_1}). This gives a nice complementarity between these two parameters, i.e., $f$ mainly probes the amplitude, and $m$ sets the peak frequency. 

\subsection{Going beyond linear calculation \label{sec:nonlin}}

The axion equation of motion~\eqref{eq:axion_eom} is at the linear order, which assumes the axion field to be homogeneous, and thus the evolution only depends on time.
However, the tachyonic production of the dark photon is momentum-dependent and thus inhomogeneous. 
The back reaction from the exponentially enhanced gauge modes could then generate inhomogeneity in the axion field through the $\phi X\tilde{X}$ term and make $\phi$ momentum dependent.
The inclusion of the ALP momentum dependence significantly complicates the equations and is therefore usually solved on the lattice~\cite{Ratzinger:2020oct}, while as a result, the GW spectrum sourced during the process may differ compared with the linear order results.

\begin{figure*}[t!]
\centering
\includegraphics[width=0.5\linewidth]{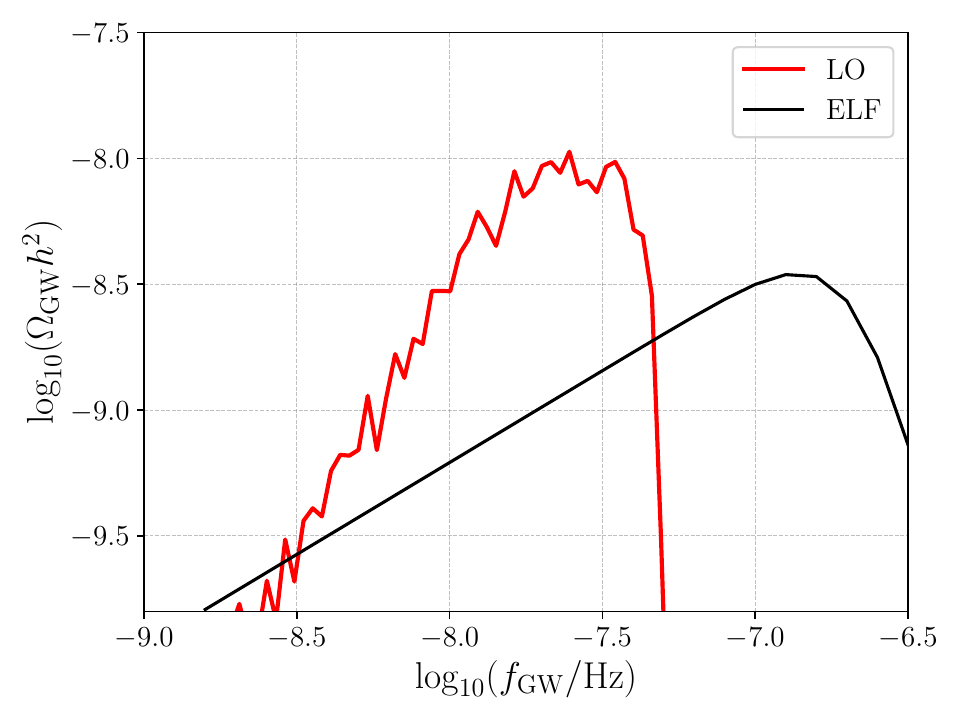}
\caption{
A comparison between the benchmark GW spectrum shown in Fig.~\ref{fig:BM} (red) and the empirical GW profile (black).
The model parameters are set to be $m\approx 8\times 10^{-12}$eV, $f\approx 8\times 10^{17}$ GeV, $\alpha=50$, the same as those in Fig.~\ref{fig:BM}.
The linear calculation gives an integrated $\Omega_{\rm GW}h^2$ about 1.5 times larger than that from the empirical fitted spectrum for this benchmark point.
}
\label{fig:GW_spectra_comparison}
\end{figure*}

From the particle point of view, the effect of the back reaction is to introduce backscattering (two dark photons into one ALP) and multiple scattering into the linear system, which shuffles energy between different modes.
The spectra of ALPs and dark photons are thus broadened.
In terms of the generated GW spectrum, these effects will lead to higher peak frequency and a more ``relaxed'' spectral shape: having a smaller power-law index at lower frequencies and dropping off slower at higher frequencies.
The amplitude of the GW spectrum, on the other hand, receives no drastic changes.
An empirical GW profile can be obtained from fitting to the lattice calculation~\cite{Madge:2023cak}, which reads
\begin{align}\label{eq:GW_empirical}
\Omega_{\rm GW}h^2&= \Omega_{\rm GW,peak}h^2\left(\dfrac{f_{\rm GW}}{f_{\rm GW,peak}}\right)^{0.73}\left(\dfrac{1}{2}\left(1+\left(\dfrac{f_{\rm GW}}{f_{\rm GW,peak}}\right)^{4.17}\right)\right)^{-1.37}\,,\\
\Omega_{\rm GW,peak}h^2&=1.2\times 10^{-7}\left(\dfrac{f}{\Mpl}\right)^4 \left(\dfrac{100}{\alpha}\right)^{2/3}\,,\nonumber\\
f_{\rm GW, peak}&=3.9\times 10^{-9}\,{\rm Hz}\left(\dfrac{\alpha}{100}\right)^{2/3}\left(\dfrac{m}{10^{-15}\ {\rm eV}}\right)^{1/2}\,.\nonumber
\end{align}
A comparison between this empirical lattice fit GW spectrum (denoted as ``ELF'' hereafter) and our linear order numeric spectrum is presented in Fig.~\ref{fig:GW_spectra_comparison}. It is important to note that this empirical expression is fitted up to $60<\alpha<100$.\footnote{Private communication with the authors of~\cite{Madge:2023cak}.}
Also, for our linear order calculation, we find that the GW signal strength drops for $\alpha<50$ due to a weakened energy transfer from the ALPs into dark photons. 
We therefore will restrict our use of the empirical fit within this range of $\alpha$.

In the next section, we will examine our linear order numeric GW spectrum as well as the empirical profile above against the NG15 observation and compare the fitting results.

\section{Numeric Calculations and Results}\label{sec:results}
\subsection{Calculation setup}\label{sec:setup}

To solve the coupled ALP-dark photon equations of motion, at the leading order we replace the friction term on the right-hand side of Eq.~\eqref{eq:axion_eom} to its expectation value, {\it i.e.} 
\begin{align}
\bv{E}\cdot\bv{B}\to\langle\bv{E}\cdot\bv{B}\rangle\nonumber=-\sum_{\lambda=\pm}\lambda \int \dfrac{k^2dk}{2\pi^2 a^4}\,{\rm Re}\left[v^\ast_\lambda(k,\tau)v^\prime_\lambda(k,\tau)\right]\,.
\end{align}
We discretize the gauge modes $v_\pm(k,\tau)$ in the wave number $k$ such that we can solve the coupled axion and dark photon equations of motion.
The range of k is chosen to be $[10^{-9}, 10^{-6}]$ Hz such that the GW signals of interest are covered (note that we have $f_{\rm GW}=k/(2\pi)$ for frequency today), which we discretize into
$N=200$ log-uniform modes.
The dark photons are assumed to be non-thermal such that all their relic abundance is generated via tachyonic production.
We then use the Bunch-Davis vacuum as the initial conditions of the gauge modes, {\it i.e.} $v_\pm(k,\tau)=e^{ik\tau}/\sqrt{2k}$.
The ALP field, on the other hand, has an initial value of $\lvert\phi_0\rvert=f$ and no initial velocity. 
Without loss of generality, we choose $\phi_0=-f$ such that we have the gauge modes $v_+$ to be more enhanced.
Furthermore, for simplicity, we ignore all the $v_-$ modes
as their energy density is very suppressed than $v_+$ modes~\cite{Machado:2018nqk,Geller:2021obo}.
The differential equations are solved from the time $\tau_{\rm min}$ where $H(\tau=\tau_{\rm min})>m$, until the time
$\tau_{\rm max}=10\tau_{\rm s}$.
We have tried various choices of $k$ range, gauge mode number $N$ and the ending time of the differential equation solution $\tau_{\rm max}$ and found the results to be consistent.

With the GW profiles numerically obtained as above, we run Markov Chain Monte-Carlo (MCMC) with {\tt emcee}~\cite{emcee} on the parameter region 
$m\in [10^{-12.5}, 10^{-10.5}]$
eV, 
$f\in[10^{17.5}, 10^{18.3}]$
GeV 
to get the posterior distribution,
with the step width chosen as ${0.1}$ and ${0.05}$ in the exponent of $10$ (logarithmic internal), respectively for the first two parameters. Whereas, for $\alpha$ direction we choose an interval of $2$ for $\alpha\in[40, 50]$ and $5$ for $\alpha\in[50, 100]$. The finer scan for the first $\alpha$ range is motivated as the GW production from this system becomes efficient around $\alpha \sim 45$.

The prior is chosen to be log-uniform for $m$ and $f$, and uniform for $\alpha$.
The likelihood is obtained from the NG15 free spectrum density shown in Ref.~\cite{NANOGrav:2023gor}, where we treat the posterior distributions in the violin plot as the likelihood distribution in excess timing delay of the corresponding frequency bin, which can be easily converted to $\log_{10}(\Omega_{\rm GW}h^2)$~\cite{Dandoy:2023jot}.
For each frequency bin, we then have a normalized likelihood function of $\log_{10}(\Omega_{\rm GW}h^2)$.
We assume there is no correlation between different frequencies, such that the overall likelihood of the GW spectra is a simple multiplication of its likelihood in each frequency bin. We have used \texttt{GetDist} to analyze the MCMC samples and generate the plots.~\cite{Lewis:2019xzd}

\subsection{Results and Constraints}
\subsubsection*{Posterior Distributions}
\begin{figure*}[t!]
\centering
\includegraphics[width=0.8\linewidth]{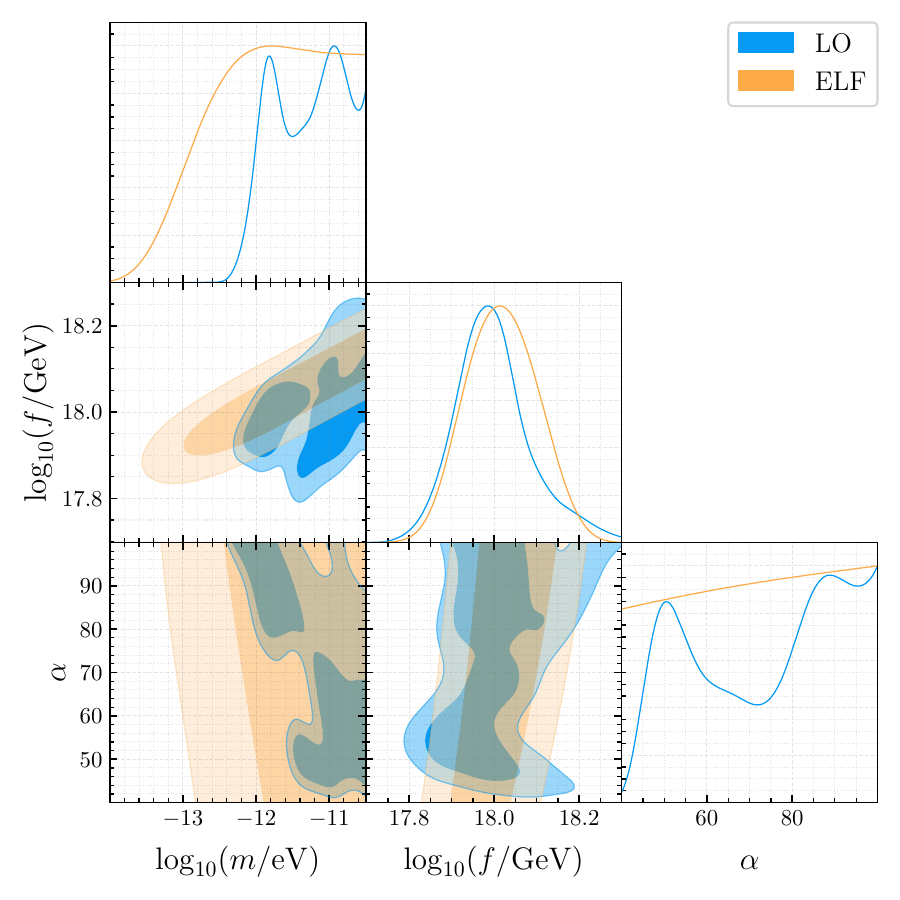}
\caption{
The posterior distribution for the parameters of the instability-induced GW model for NG15 observation. 
The darker and lighter regions are for the 68\% and 95\% CL regions. The blue and orange shaded areas correspond to our linear order (LO) numeric GW spectra and the empiric GW spectra for the lattice calculation~\cite{Madge:2023cak} (empirical lattice fit, ELF), respectively.
}
\label{fig:MCMC}
\end{figure*}

In this section, we present the results of the MCMC analysis for the parameter regions favored by the NG15 observation and derive the constraints on this scenario. The favored regions are shown for the linear (blue) and lattice fitted analysis (orange) in Fig.~\ref{fig:MCMC} where we plot the posterior distribution triangle plots. We first see that the two analyses differ by an order of magnitude in the 1d posterior of the ALP mass, with masses above $10^{-13}-10^{-12}$ preferred. The ALP mass mostly determines the peak frequency, see Eq.~\eqref{eq:GW_peak}, which changes between the two analyses by around an order of magnitude. In contrast, the preferred $f$ regions are very similar -- as expected from the similar GW magnitudes in the two analyses (recall that $\Omega_{\rm GW}\propto f^4$ as derived at the end of Sec.~\ref{sec:model}). In both cases, the favored $f$ values lie around $8\times 10^{17}$ GeV, almost independently of the other parameters. 

We do not find a strong dependence on $\alpha$, however, two comments are in order. Firstly, for small $\alpha\lesssim 50$ signal from the LO calculation disappears, and so the plot cannot be extended into the small $\alpha$ region. For large $\alpha$, we cannot trust the linear order (LO) calculation due to the large back reaction, and the empirical lattice fit Eq.~\eqref{eq:GW_empirical} is only valid for $\alpha<100$ as explained in Sec.~\ref{sec:nonlin}.
The higher $\alpha$ region is therefore the only loophole to our conclusions here.

Comparing with the results of~\cite{Ratzinger:2020oct,Madge:2023cak} for the same model, but with the previous data, we find that the preferred values of $m$ and $f$ by the NG15 data are significantly higher than those preferred by NG12.
This can be explained by the inclusion of higher frequency bins in the NG15 analysis~\cite{NANOGrav:2023gor} compared with the NG12~\cite{NANOGrav:2020bcs} data, and moreover, the signal strengths in these bins imply GWs of a higher magnitude. As a result, both the peak frequency and the inferred magnitude are higher, resulting in higher $m$ and $f$ (see Eq.~\eqref{eq:GW_peak}). Following the same argument we can infer the fitting results on data from the other PTAs, {\it i.e.}, EPTA, and PPTA. The data used for SGWB analysis in EPTA and PPTA extends to even larger GW frequencies and spectral amplitudes than NG15 and therefore would favor an even larger $f$ and $m$. See Appendix~\ref{app:EPTA} for some more details.

\subsubsection*{Cosmological Constraints}
We now move to discuss the constraints on this scenario from cosmological observations, mainly arising from the energy density in the ALP and gauge fields. After the tachyonic production stops being efficient, the ALPs start to behave like matter.
Given the axion-dark photon coupling, the decay width of $\phi\to XX$ can be shown to be $\Gamma_{\phi\to XX}=\alpha^2 m^3/(64\pi f^2)$, implying the ALPs to be cosmologically long-lived in the parameter space of interest.
The model is therefore excluded if the dark matter abundance is higher than the observed one ({\it i.e.} over-closure) and if the dark radiation exceeds the bounds on $\Delta N_{\rm eff}$. 
The energy densities of the ALPs and the dark photons are calculated as
\begin{align}
\rho_\phi&=\dfrac{1}{2a^2}\phi^{\prime 2}+\dfrac{1}{2}m^2\phi^2\,,\\
\rho_X &= \dfrac{1}{2a^4}\int\dfrac{d^3k}{(2\pi)^3}\left(\left(\lvert v^\prime_+\rvert^2 + \lvert v^\prime_-\rvert^2\right)+k^2\left(\lvert v_+\rvert^2 + \lvert v_-\rvert^2\right)-2k\right)\,,
\end{align}
which we evaluate at the time when the tachyonic production finishes and ALPs and dark photons start to behave like normal matter and radiation, and hence can be redshifted to calculate the fraction of their energy density today.

\begin{figure*}[h!]
\centering
\includegraphics[width=0.49\linewidth]{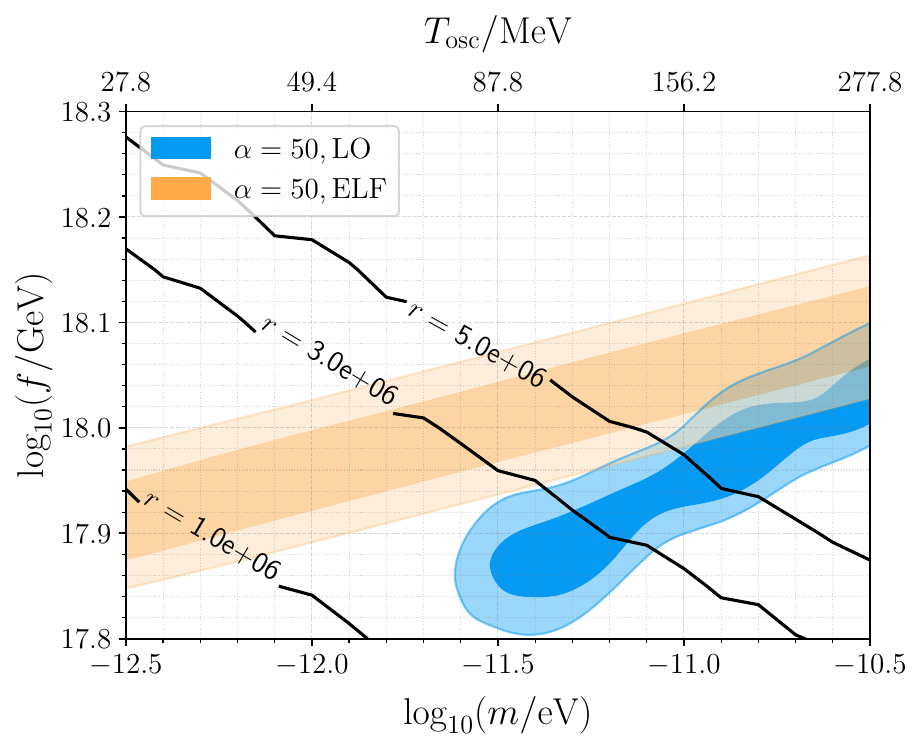}
\includegraphics[width=0.49\linewidth]{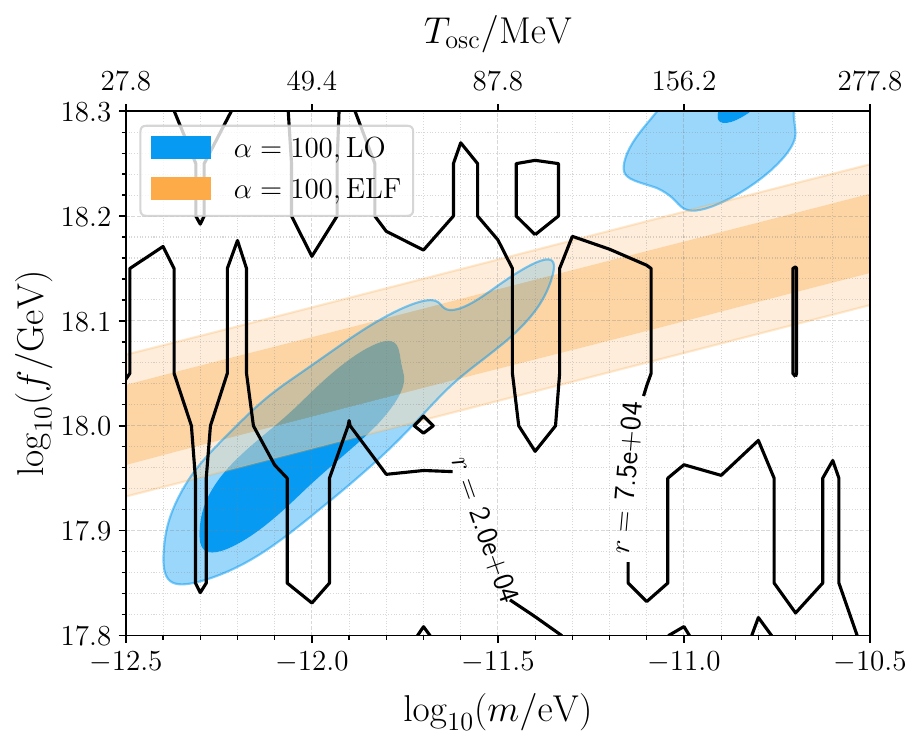} 
\caption{The $r\equiv\rho_\phi/\rho_{\rm DM}$ contours of various values shown on top of the posterior distribution of fixed $\alpha$, where the $\alpha$ values are chosen to be 50 and 100 for the left and right panel, respectively. 
The plot range is chosen to focus on the favored region of the linear order calculations, and a more complete distribution of the empirical spectra can be found in Fig.~\ref{fig:Neff}.
The ratio $r$ is obtained from the linear order calculations.
The NG15 favored mass regions have the axion rolling and tachyonic production before BBN, as can be seen from the axion oscillation temperature $T_{\rm osc}$ (the cosmic temperature where $H=m$) shown on the top axes.
In both panels, the NG15 preferred parameter space will generate too large an $r$ and thus overclose the universe, and we have checked the situation to be the same for the other values of $\alpha$.
}
\label{fig:overclosure}
\end{figure*}

The issue of over-closure was already pointed out in ~\cite{Kitajima:2017peg,Machado:2019xuc,Madge:2023cak}, and we find similarly that the relic abundance of the ALP exceeds that of the observed DM abundance by more than 4 orders of magnitude in all the preferred parameter space.
See Fig.~\ref{fig:overclosure} where we show the values of the ALP to DM density ratio $r$ for $\alpha=50,100$ in the linear calculation and both the linear and lattice-fitted posterior analyses. In drawing the constant $r$ contours in the figure, we have averaged over ALP energy density from $\tau_{\rm max}/4 $ to $\tau_{\rm max} $ by properly taking into account the redshift dependence. This results in smoother contours and the averaging period is well beyond the end of the tachyonic band which occurs at $\tau \approx {\tau_{\rm max}/10}$ signifying the end of particle production.
The issue of overproduction of DM may be generic on ALP models if they are to be the major source of nanohertz SGWB~\cite{Eroncel:2022vjg}.
There have been some proposals to mitigate the over-closure problem with extra model building that tries to suppress the axion mass (hence energy density) at late times. The models exhibit dilution of the axion energy density at late times which can be achieved via a time-dependent axion potential originating from dark sector phase transition, monopole effects, etc~\cite{Ratzinger:2020oct,Namba:2020kij,Heurtier:2021rko}. The NG15 data favors a higher mass of axion, hence a higher $T_{\rm osc} \approx \mathcal{O}(10 -100)~{\rm MeV}$. Thus it leaves a wider window in temperature to model away the overdensity problem before the matter-radiation equality at $T\approx 1 {\rm eV}$.

\begin{figure*}[t!]
\centering
\includegraphics[width=0.49\linewidth]{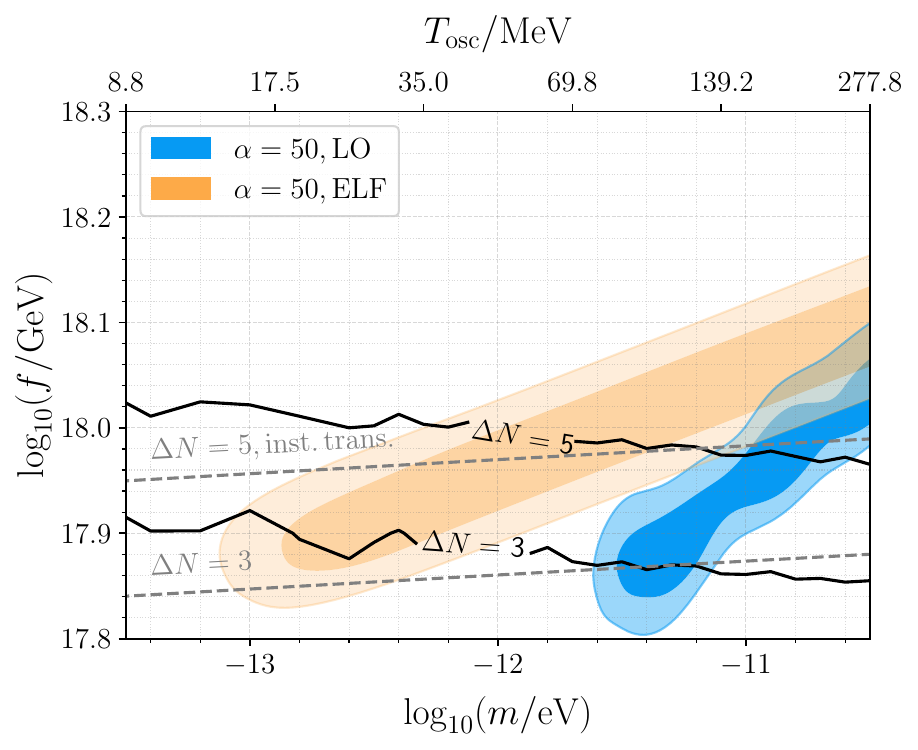}
\includegraphics[width=0.49\linewidth]{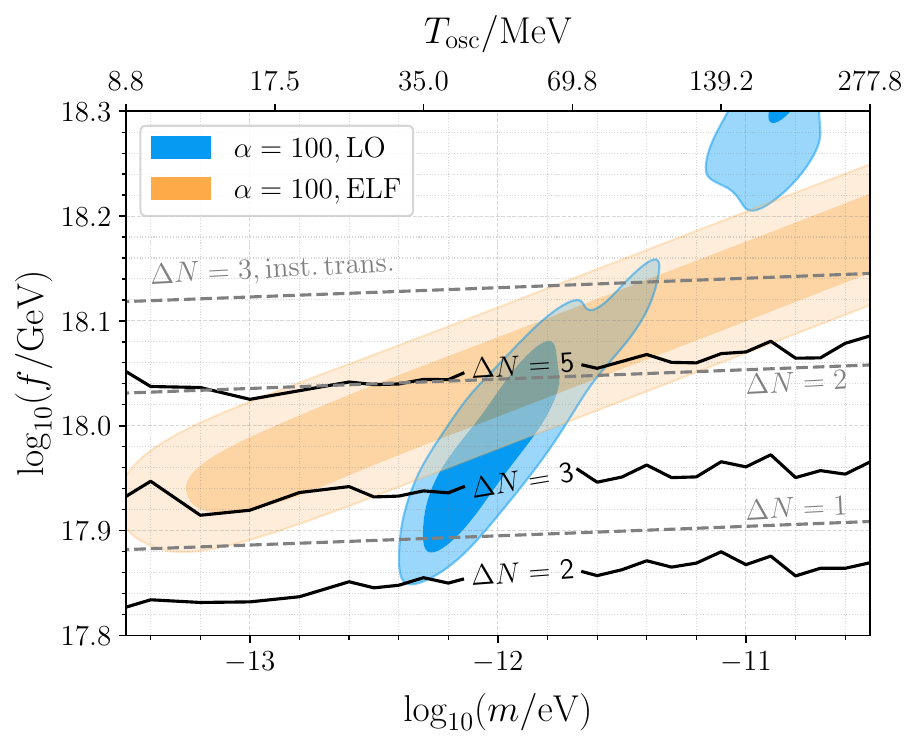} 
\caption{
Similar to Fig.~\ref{fig:overclosure}, but the solid contours are for $\Delta N_{\rm eff}$ obtained in the linear calculations.
The plot range is zoomed out to include a more complete parameter space of the empirical spectral.
$\Delta N$ in the plot is short for $\Delta N_{\rm eff}$.
The dashed gray lines are the $\Delta N_{\rm eff}$ contours obtained by assuming instantaneous energy transfer from the ALP to the dark photons~\cite{Ratzinger:2020oct}.
Models with parameters favored by NG15 severely violate the current constraints on $\Delta N_{\rm eff}$.
}
\label{fig:Neff}
\end{figure*}

We now move to the $\Delta N_{\rm eff}$ constraint.
The current measurement on the CMB requires $\Delta N_{\rm eff}\lesssim 0.3$~\cite{Planck:2018vyg}, and BBN gives a slightly weaker constraint of $\Delta N_{\rm eff}\lesssim 0.4$~\cite{Adshead:2020htj}.
In Fig.~\ref{fig:Neff} we show the posterior distribution of $m$ and $f$ for $\alpha=50$ and 100 together with the contours of $\Delta N_{\rm eff}$. The black $\Delta N_{\rm eff}$ contours are obtained from linear order calculation. Similar to the $r$ contours we have employed similar averaging of DR energy densities for smoother contours.
We also show the results for the $\Delta N_{\rm eff}$ calculation (in dashed lines) from Ref.~\cite{Ratzinger:2020oct,Madge:2023cak}, where the authors assumed instantaneous deposition of the total energy from ALP to DR at scale factor $a_*$ where $\rho_{\rm ALP} \approx \rho_{\rm DR}$. 
This approximation gives similar results to our LO $\Delta N_{\rm eff}$ calculation for small $\alpha$.
For higher values of $\alpha$, in our linear calculation, not all the ALP energy is transferred to the dark photons at $a=a_\ast$, and a substantial portion is released gradually in the later evolution. $\Delta N_{\rm eff}$ is then expected to increase
by this process as the residual $\Omega_{\rm ALP}$ 
keeps growing $(\propto a)$ untill it decays to $\Omega_{\rm DR}$. We find that our linear result is higher than the one from the instantaneous assumption by a factor of $2-2.5$, as we can see on the right side of Fig.~\ref{fig:Neff}. We stress that for these values of $\alpha$, the lattice calculation can significantly alter the dynamics and we do not know whether this effect persists in the lattice. Therefore, we will show both results and treat the instantaneous assumption as a conservative estimate of the $\Delta N_{\rm eff}$ as it can only underestimate it. A further lattice study is required to confirm our results here. 
We see that for the preferred region, in both the linear and the lattice fitted analyses, $\Delta N_{\rm eff} \gtrsim 1-3$, significantly above the current constraint.

Another way to understand the $\Delta N_{\rm eff} $ constraint is to isolate the contributions from the background DR and GW spectrum. For our linear calculations, we find that across all the viable parameter spaces, the $\Delta N_{\rm eff}$ contribution from GWs $(\Delta N_{\rm eff,GW})$ is related to the $\Delta N_{\rm eff}$ of DR $(\Delta N_{\rm eff,DR})$ by
\begin{equation}
    \label{eq:neffrel}
    \frac{\Delta N_{\rm eff,DR}}{\left(\Delta N_{\rm eff,GW}\right)^{\frac{1}{2}} }\approx 70\left(30\right)\;,
\end{equation}
where $\Delta N_{\rm eff,GW} $ is calculated as $\int (d\Omega_{\rm GW} / d\ln k) (dk / k)$. The value inside the parenthesis corresponds to the instantaneous assumption with $\alpha$ close to 100. This approximate scaling relation holds for all parameter points and for the $\alpha$ range $[50,100]$ used in this paper. This is because the GW spectrum is the two-point power spectrum for the tensor fluctuation in the DR and is thus related to the square of DR energy density. For the GW power spectrum to explain the NG15 observation (like the benchmark on Fig.~\ref{fig:BM}), it would require $\Delta N_{\rm eff,GW} \approx 0.003$. 
This in turn gives $\Delta N_{\rm eff,DR} \approx 1-4$ using Eq.~\eqref{eq:neffrel} and thus rules out the parameter point from the cosmological $\Delta N_{\rm eff}$ measurement. 
Conventional mechanisms that dilute $\Delta N_{\rm eff}$ through entropy injection will not work since that will also dilute the GW signal by the same fraction. From Eq.~\eqref{eq:neffrel} it naively seems that one can decrease their ratio $\Delta N_{\rm eff,DR} / \Delta N_{\rm eff,GW}$ by increasing both $\rho_{\rm GW}$ and $\rho_{\rm DR}$ while holding $\rho_{\rm DR}/\rho^{1/2}_{\rm GW}$ fixed at the end of tachyonic production. However, any such significant increase would result in the dark sector dominating the energy density of the universe, where we don't expect the scaling in Eq.~\eqref{eq:neffrel} to hold. The analysis of this scenario is beyond the scope of this work. 

Given the tight cosmological constraints from the dark matter relic abundance and $\Delta N_{\rm eff}$, one may inspect the maximum SGWB amplitude that could be generated by the tachyonic instability mechanism. 
If the over-abundant axions can be depleted with additional model building, we will need to decrease $\Delta N_{\rm eff,DR}$ by an order of magnitude to be compatible with the current CMB measurements, which renders the GW amplitude to reduce by about two orders of magnitude, making up about 1\% of the measured NG15 signal.
However, if it turns out that the over-closure bound cannot be modeled away, the current setup will imply the induced GW can only correspond to a minuscule $<\mathcal{O}(10^{-8} - 10^{-12})$ fraction of the observed GW spectrum depending on the $\alpha$ values.

\section{Conclusion and Discussion}
In this work, we explore the possibility that the reported SGWB found in the NANOGrav 15-year data set~\cite{NANOGrav:2023gor} is generated from the coupled dynamics between the ALPs and the dark photons.
The ALPs' rolling exponentially amplifies the massless gauge modes, and the drastic change of the gauge boson mode functions then sources GWs which
can potentially explain the observed SGWB spectrum.
However, as we have shown in this paper, the parameter region that can explain the NG15 observation is incompatible with the observed dark matter relic abundance in the current universe, as well as the $\Delta N_{\rm eff}$ constraints from BBN and CMB.

We have additionally checked the robustness of our results. We have carried out numerical calculations at the linear order, and have compared with the results from lattice-fitted formulae from~\cite{Madge:2023cak}. We find that the results remain independent. We do not, however, explore the large back-reaction parameter region with $\alpha>100$ for which both analyses are invalid and new lattice calculations are required, so our argument may not apply in that region. 
A naive extension of our fit into this region, however, would only strengthen the constraints due to the positive correlation between $\alpha$ and $f$ in the GW peak position.

When generating the MCMC chains, we use the posterior ``violins'' of the NG15 free spectrum density as the likelihood function of the sample points.
It is known that this statistics approach usually results in a slightly smaller favored GW amplitude compared with going through the whole pipeline, as it ignores the correlation between different frequency components~\cite{Bringmann:2023opz}.
The tension between the instability-induced GW explanation of NG15 and the cosmological constraints from dark matter relic abundance and $\Delta N_{\rm eff}$ could therefore be even stronger than what we have pointed out in the main text.
Thus, a more sophisticated analysis most likely will strengthen the bounds and will not change the conclusion of the paper.

The stringent constraints on the vanilla ALP scenario warrant additional model building to address the over-closure and $\Delta N_{\rm eff}$ constraint. Although several mechanisms exist to address the ALP over-closure, reducing $\Delta N_{\rm eff}$ without comprising the GW signal is highly challenging. Thus the NG15 observation in conjunction with other cosmological measurements strongly rules out the instability-induced GW from the vanilla ALP model to be the entire signal in the $ m \approx 10^{-14} - 10^{-11}$ eV ALP mass range. It can still potentially be an at most percent-level subdominant signal if another source of GW, such as supermassive black hole binaries, is responsible for the NG15 signal, and it may be of particular interest to study how such a signal can be potentially probed.

\section*{Acknowledgement}
We thank Yang Bai, Chao Chen, and Kai-Feng Zheng for useful discussions. We also thank Wolfram Ratzinger and Eric Madge for clarifications about \cite{Madge:2023cak}.
MG is supported in part by the Israel Science Foundation under Grant No. 1302/19.  MG is also supported in part by the US-Israeli BSF grant 2018236 and the NSF-BSF grant 2021779. SG and YT are supported by the U.S. National
Science Foundation grant PHY2112540.
SL is supported by the Area of Excellence (AoE) under the Grant No. AoE/P-404/18-3 issued by the Research Grants Council of Hong Kong S.A.R. This work was prepared in part at the Aspen Center for Physics, which is supported by National Science Foundation grant PHY-2210452.

\appendix
\section{Instability-induced GW on the other PTAs}
\label{app:EPTA}

\begin{figure}[t]
    \centering
    \includegraphics[width=0.48\textwidth]{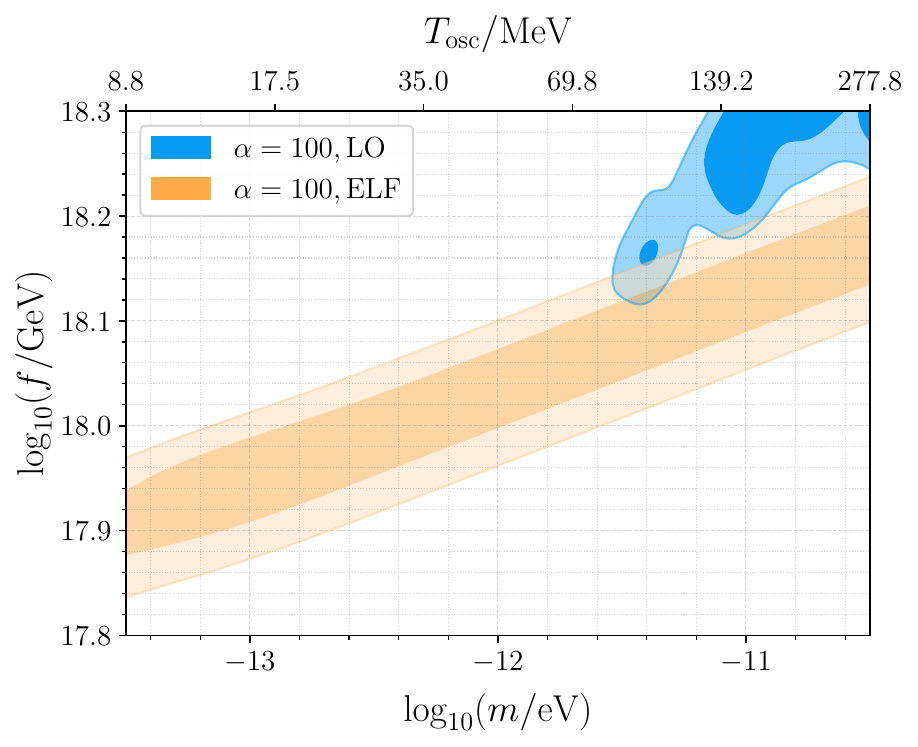} \hspace{2mm}
    \includegraphics[width=0.48\textwidth]{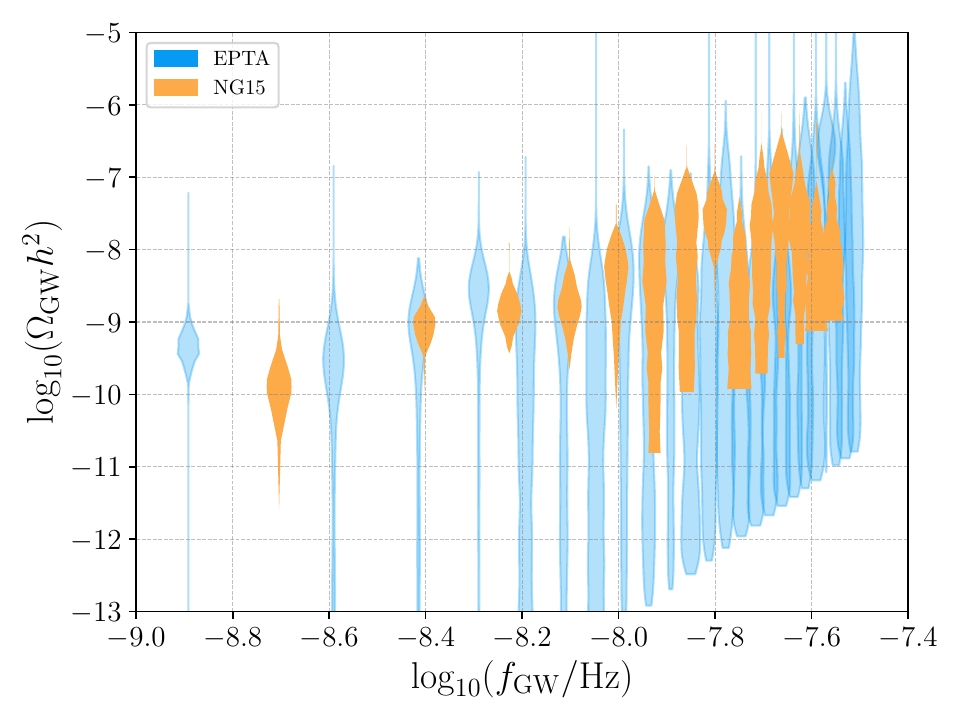}
    \caption{{\it Left:} Our fitted result to the EPTA released HD free spectrum~\cite{EPTA:2023sfo} for fixed $\alpha=100$, companion to the right panels of Fig. 4 and 5. {\it Right}: Comparison between the posterior ``violins'' used for the two fittings.}
    \label{fig:fit_EPTA}
\end{figure}

In principle, our analysis can be performed on the other PTA results and see if our conclusion changes. Here we use EPTA as an example, repeating our fitting process on their released chain file of the HD-correlated free spectrum~\cite{EPTA:2023sfo} (using the ``DR2full'' dataset).
The result for $\alpha=100$ is shown in the left panel of Fig.~\ref{fig:fit_EPTA}, to be compared with the right panels of Fig.~\ref{fig:Neff}.
The fitting result favors an even larger axion constant $f$ and axion mass $m$, indicating a worse consistency with the other cosmological constraints.
This can be explained by comparing the posterior ``violins'' used in the two fittings, as shown in the right panel of Fig.~\ref{fig:fit_EPTA}.
The EPTA data use 24 frequency bins~\cite{EPTA:2023sfo}, running to a larger GW frequency as well as $\Omega_{\rm GW}h^2$ compared with the NG15 data, and therefore result in larger $f$ and $m$.
For the case of PPTA, a total of 28 frequency bins are involved, which extends to even larger frequency and amplitude~\cite{Reardon:2023gzh}. Therefore, following the same argument, PPTA cannot improve our situation either.

\bibliography{b-mode}

\providecommand{\href}[2]{#2}\begingroup\raggedright\begin{thebibliography}{10}

\bibitem{NANOGrav:2023gor}
{\bf NANOGrav} Collaboration, G.~Agazie et~al., {\it {The NANOGrav 15-year Data
  Set: Evidence for a Gravitational-Wave Background}},
  \href{http://arxiv.org/abs/2306.16213}{{\tt arXiv:2306.16213}}.

\bibitem{Xu:2023wog}
H.~Xu et~al., {\it {Searching for the nano-Hertz stochastic gravitational wave
  background with the Chinese Pulsar Timing Array Data Release I}},
  \href{http://arxiv.org/abs/2306.16216}{{\tt arXiv:2306.16216}}.

\bibitem{Antoniadis:2023aac}
J.~Antoniadis et~al., {\it {The second data release from the European Pulsar
  Timing Array IV. Search for continuous gravitational wave signals}},
  \href{http://arxiv.org/abs/2306.16226}{{\tt arXiv:2306.16226}}.

\bibitem{Reardon:2023gzh}
D.~J. Reardon et~al., {\it {Search for an isotropic gravitational-wave
  background with the Parkes Pulsar Timing Array}},
  \href{http://arxiv.org/abs/2306.16215}{{\tt arXiv:2306.16215}}.

\bibitem{Phinney:2001di}
E.~S. Phinney, {\it {A Practical theorem on gravitational wave backgrounds}},
  \href{http://arxiv.org/abs/astro-ph/0108028}{{\tt astro-ph/0108028}}.

\bibitem{Sesana:2012ak}
A.~Sesana, {\it {Systematic investigation of the expected gravitational wave
  signal from supermassive black hole binaries in the pulsar timing band}},
  {\em Mon. Not. Roy. Astron. Soc.} {\bf 433} (2013) 1,
  [\href{http://arxiv.org/abs/1211.5375}{{\tt arXiv:1211.5375}}].

\bibitem{Chen:2018rzo}
Z.-C. Chen, F.~Huang, and Q.-G. Huang, {\it {Stochastic Gravitational-wave
  Background from Binary Black Holes and Binary Neutron Stars and Implications
  for LISA}},  {\em Astrophys. J.} {\bf 871} (2019), no.~1 97,
  [\href{http://arxiv.org/abs/1809.10360}{{\tt arXiv:1809.10360}}].

\bibitem{Zhu:2011bd}
X.-J. Zhu, E.~Howell, T.~Regimbau, D.~Blair, and Z.-H. Zhu, {\it {Stochastic
  Gravitational Wave Background from Coalescing Binary Black Holes}},  {\em
  Astrophys. J.} {\bf 739} (2011) 86,
  [\href{http://arxiv.org/abs/1104.3565}{{\tt arXiv:1104.3565}}].

\bibitem{Freese:2022qrl}
K.~Freese and M.~W. Winkler, {\it {Have pulsar timing arrays detected the hot
  big bang: Gravitational waves from strong first order phase transitions in
  the early Universe}},  {\em Phys. Rev. D} {\bf 106} (2022), no.~10 103523,
  [\href{http://arxiv.org/abs/2208.03330}{{\tt arXiv:2208.03330}}].

\bibitem{Levi:2022bzt}
N.~Levi, T.~Opferkuch, and D.~Redigolo, {\it {The supercooling window at weak
  and strong coupling}},  {\em JHEP} {\bf 02} (2023) 125,
  [\href{http://arxiv.org/abs/2212.08085}{{\tt arXiv:2212.08085}}].

\bibitem{Ellis:2020nnr}
J.~Ellis, M.~Lewicki, and V.~Vaskonen, {\it {Updated predictions for
  gravitational waves produced in a strongly supercooled phase transition}},
  {\em JCAP} {\bf 11} (2020) 020, [\href{http://arxiv.org/abs/2007.15586}{{\tt
  arXiv:2007.15586}}].

\bibitem{Lewicki:2020azd}
M.~Lewicki and V.~Vaskonen, {\it {Gravitational waves from colliding vacuum
  bubbles in gauge theories}},  {\em Eur. Phys. J. C} {\bf 81} (2021), no.~5
  437, [\href{http://arxiv.org/abs/2012.07826}{{\tt arXiv:2012.07826}}].
  [Erratum: Eur.Phys.J.C 81, 1077 (2021)].

\bibitem{Lewicki:2020jiv}
M.~Lewicki and V.~Vaskonen, {\it {Gravitational wave spectra from strongly
  supercooled phase transitions}},  {\em Eur. Phys. J. C} {\bf 80} (2020),
  no.~11 1003, [\href{http://arxiv.org/abs/2007.04967}{{\tt
  arXiv:2007.04967}}].

\bibitem{Hindmarsh:2019phv}
M.~Hindmarsh and M.~Hijazi, {\it {Gravitational waves from first order
  cosmological phase transitions in the Sound Shell Model}},  {\em JCAP} {\bf
  12} (2019) 062, [\href{http://arxiv.org/abs/1909.10040}{{\tt
  arXiv:1909.10040}}].

\bibitem{Caprini:2019egz}
C.~Caprini et~al., {\it {Detecting gravitational waves from cosmological phase
  transitions with LISA: an update}},  {\em JCAP} {\bf 03} (2020) 024,
  [\href{http://arxiv.org/abs/1910.13125}{{\tt arXiv:1910.13125}}].

\bibitem{Zu:2023olm}
L.~Zu, C.~Zhang, Y.-Y. Li, Y.-C. Gu, Y.-L.~S. Tsai, and Y.-Z. Fan, {\it {Mirror
  QCD phase transition as the origin of the nanohertz Stochastic
  Gravitational-Wave Background detected by the Pulsar Timing Arrays}},
  \href{http://arxiv.org/abs/2306.16769}{{\tt arXiv:2306.16769}}.

\bibitem{Franciolini:2023wjm}
G.~Franciolini, D.~Racco, and F.~Rompineve, {\it {Footprints of the QCD
  Crossover on Cosmological Gravitational Waves at Pulsar Timing Arrays}},
  \href{http://arxiv.org/abs/2306.17136}{{\tt arXiv:2306.17136}}.

\bibitem{Megias:2023kiy}
E.~Megias, G.~Nardini, and M.~Quiros, {\it {Pulsar Timing Array Stochastic
  Background from light Kaluza-Klein resonances}},
  \href{http://arxiv.org/abs/2306.17071}{{\tt arXiv:2306.17071}}.

\bibitem{Han:2023olf}
C.~Han, K.-P. Xie, J.~M. Yang, and M.~Zhang, {\it {Self-interacting dark matter
  implied by nano-Hertz gravitational waves}},
  \href{http://arxiv.org/abs/2306.16966}{{\tt arXiv:2306.16966}}.

\bibitem{Fujikura:2023lkn}
K.~Fujikura, S.~Girmohanta, Y.~Nakai, and M.~Suzuki, {\it {NANOGrav Signal from
  a Dark Conformal Phase Transition}},
  \href{http://arxiv.org/abs/2306.17086}{{\tt arXiv:2306.17086}}.

\bibitem{Li:2023bxy}
S.-P. Li and K.-P. Xie, {\it {A collider test of nano-Hertz gravitational waves
  from pulsar timing arrays}},  \href{http://arxiv.org/abs/2307.01086}{{\tt
  arXiv:2307.01086}}.

\bibitem{Ashoorioon:2022raz}
A.~Ashoorioon, K.~Rezazadeh, and A.~Rostami, {\it {NANOGrav signal from the end
  of inflation and the LIGO mass and heavier primordial black holes}},  {\em
  Phys. Lett. B} {\bf 835} (2022) 137542,
  [\href{http://arxiv.org/abs/2202.01131}{{\tt arXiv:2202.01131}}].

\bibitem{Athron:2023mer}
P.~Athron, A.~Fowlie, C.-T. Lu, L.~Morris, L.~Wu, Y.~Wu, and Z.~Xu, {\it {Can
  Supercooled Phase Transitions explain the Gravitational Wave Background
  observed by Pulsar Timing Arrays?}},
  \href{http://arxiv.org/abs/2306.17239}{{\tt arXiv:2306.17239}}.

\bibitem{Ghosh:2023aum}
T.~Ghosh, A.~Ghoshal, H.-K. Guo, F.~Hajkarim, S.~F. King, K.~Sinha, X.~Wang,
  and G.~White, {\it {Did we hear the sound of the Universe boiling? Analysis
  using the full fluid velocity profiles and NANOGrav 15-year data}},
  \href{http://arxiv.org/abs/2307.02259}{{\tt arXiv:2307.02259}}.

\bibitem{Samanta:2020cdk}
R.~Samanta and S.~Datta, {\it {Gravitational wave complementarity and impact of
  NANOGrav data on gravitational leptogenesis}},  {\em JHEP} {\bf 05} (2021)
  211, [\href{http://arxiv.org/abs/2009.13452}{{\tt arXiv:2009.13452}}].

\bibitem{DiBari:2023upq}
P.~Di~Bari and M.~H. Rahat, {\it {The split majoron model confronts the
  NANOGrav signal}},  \href{http://arxiv.org/abs/2307.03184}{{\tt
  arXiv:2307.03184}}.

\bibitem{Auclair:2019wcv}
P.~Auclair et~al., {\it {Probing the gravitational wave background from cosmic
  strings with LISA}},  {\em JCAP} {\bf 04} (2020) 034,
  [\href{http://arxiv.org/abs/1909.00819}{{\tt arXiv:1909.00819}}].

\bibitem{Blasi:2020mfx}
S.~Blasi, V.~Brdar, and K.~Schmitz, {\it {Has NANOGrav found first evidence for
  cosmic strings?}},  {\em Phys. Rev. Lett.} {\bf 126} (2021), no.~4 041305,
  [\href{http://arxiv.org/abs/2009.06607}{{\tt arXiv:2009.06607}}].

\bibitem{Ellis:2020ena}
J.~Ellis and M.~Lewicki, {\it {Cosmic String Interpretation of NANOGrav Pulsar
  Timing Data}},  {\em Phys. Rev. Lett.} {\bf 126} (2021), no.~4 041304,
  [\href{http://arxiv.org/abs/2009.06555}{{\tt arXiv:2009.06555}}].

\bibitem{Ferrer:2023uwz}
F.~Ferrer, A.~Ghoshal, and M.~Lewicki, {\it {Imprints of a Supercooled Universe
  in the Gravitational Wave Spectrum from a Cosmic String network}},
  \href{http://arxiv.org/abs/2304.02636}{{\tt arXiv:2304.02636}}.

\bibitem{Wang:2023len}
Z.~Wang, L.~Lei, H.~Jiao, L.~Feng, and Y.-Z. Fan, {\it {The nanohertz
  stochastic gravitational-wave background from cosmic string Loops and the
  abundant high redshift massive galaxies}},
  \href{http://arxiv.org/abs/2306.17150}{{\tt arXiv:2306.17150}}.

\bibitem{Ellis:2023tsl}
J.~Ellis, M.~Lewicki, C.~Lin, and V.~Vaskonen, {\it {Cosmic Superstrings
  Revisited in Light of NANOGrav 15-Year Data}},
  \href{http://arxiv.org/abs/2306.17147}{{\tt arXiv:2306.17147}}.

\bibitem{Saikawa:2017hiv}
K.~Saikawa, {\it {A review of gravitational waves from cosmic domain walls}},
  {\em Universe} {\bf 3} (2017), no.~2 40,
  [\href{http://arxiv.org/abs/1703.02576}{{\tt arXiv:1703.02576}}].

\bibitem{Ferreira:2022zzo}
R.~Z. Ferreira, A.~Notari, O.~Pujolas, and F.~Rompineve, {\it {Gravitational
  waves from domain walls in Pulsar Timing Array datasets}},  {\em JCAP} {\bf
  02} (2023) 001, [\href{http://arxiv.org/abs/2204.04228}{{\tt
  arXiv:2204.04228}}].

\bibitem{Kitajima:2023cek}
N.~Kitajima, J.~Lee, K.~Murai, F.~Takahashi, and W.~Yin, {\it {Nanohertz
  Gravitational Waves from Axion Domain Walls Coupled to QCD}},
  \href{http://arxiv.org/abs/2306.17146}{{\tt arXiv:2306.17146}}.

\bibitem{Bai:2023cqj}
Y.~Bai, T.-K. Chen, and M.~Korwar, {\it {QCD-Collapsed Domain Walls: QCD Phase
  Transition and Gravitational Wave Spectroscopy}},
  \href{http://arxiv.org/abs/2306.17160}{{\tt arXiv:2306.17160}}.

\bibitem{Lazarides:2023ksx}
G.~Lazarides, R.~Maji, and Q.~Shafi, {\it {Superheavy quasi-stable strings and
  walls bounded by strings in the light of NANOGrav 15 year data}},
  \href{http://arxiv.org/abs/2306.17788}{{\tt arXiv:2306.17788}}.

\bibitem{Barman:2023fad}
B.~Barman, D.~Borah, S.~Jyoti~Das, and I.~Saha, {\it {Scale of Dirac
  leptogenesis and left-right symmetry in the light of recent PTA results}},
  \href{http://arxiv.org/abs/2307.00656}{{\tt arXiv:2307.00656}}.

\bibitem{Sakharov:2021dim}
A.~S. Sakharov, Y.~N. Eroshenko, and S.~G. Rubin, {\it {Looking at the NANOGrav
  signal through the anthropic window of axionlike particles}},  {\em Phys.
  Rev. D} {\bf 104} (2021), no.~4 043005,
  [\href{http://arxiv.org/abs/2104.08750}{{\tt arXiv:2104.08750}}].

\bibitem{King:2023cgv}
S.~F. King, D.~Marfatia, and M.~H. Rahat, {\it {Towards distinguishing Dirac
  from Majorana neutrino mass with gravitational waves}},
  \href{http://arxiv.org/abs/2306.05389}{{\tt arXiv:2306.05389}}.

\bibitem{Ge:2023rce}
S.~Ge, {\it {Stochastic gravitational wave background: birth from axionic
  string-wall death}},  \href{http://arxiv.org/abs/2307.08185}{{\tt
  arXiv:2307.08185}}.

\bibitem{Pi:2020otn}
S.~Pi and M.~Sasaki, {\it {Gravitational Waves Induced by Scalar Perturbations
  with a Lognormal Peak}},  {\em JCAP} {\bf 09} (2020) 037,
  [\href{http://arxiv.org/abs/2005.12306}{{\tt arXiv:2005.12306}}].

\bibitem{Baumann:2007zm}
D.~Baumann, P.~J. Steinhardt, K.~Takahashi, and K.~Ichiki, {\it {Gravitational
  Wave Spectrum Induced by Primordial Scalar Perturbations}},  {\em Phys. Rev.
  D} {\bf 76} (2007) 084019, [\href{http://arxiv.org/abs/hep-th/0703290}{{\tt
  hep-th/0703290}}].

\bibitem{Dandoy:2023jot}
V.~Dandoy, V.~Domcke, and F.~Rompineve, {\it {Search for scalar induced
  gravitational waves in the International Pulsar Timing Array Data Release 2
  and NANOgrav 12.5 years dataset}},
  \href{http://arxiv.org/abs/2302.07901}{{\tt arXiv:2302.07901}}.

\bibitem{Choudhury:2023rks}
S.~Choudhury, S.~Panda, and M.~Sami, {\it {Quantum loop effects on the power
  spectrum and constraints on primordial black holes}},
  \href{http://arxiv.org/abs/2303.06066}{{\tt arXiv:2303.06066}}.

\bibitem{Cai:2023dls}
Y.-F. Cai, X.-C. He, X.~Ma, S.-F. Yan, and G.-W. Yuan, {\it {Limits on
  scalar-induced gravitational waves from the stochastic background by pulsar
  timing array observations}},  \href{http://arxiv.org/abs/2306.17822}{{\tt
  arXiv:2306.17822}}.

\bibitem{Lozanov:2023aez}
K.~D. Lozanov, M.~Sasaki, and V.~Takhistov, {\it {Universal Gravitational Wave
  Signatures of Cosmological Solitons}},
  \href{http://arxiv.org/abs/2304.06709}{{\tt arXiv:2304.06709}}.

\bibitem{Lozanov:2022yoy}
K.~D. Lozanov and V.~Takhistov, {\it {Enhanced Gravitational Waves from
  Inflaton Oscillons}},  {\em Phys. Rev. Lett.} {\bf 130} (2023), no.~18
  181002, [\href{http://arxiv.org/abs/2204.07152}{{\tt arXiv:2204.07152}}].

\bibitem{Broadhurst:2023tus}
T.~Broadhurst, C.~Chen, T.~Liu, and K.-F. Zheng, {\it {Binary Supermassive
  Black Holes Orbiting Dark Matter Solitons: From the Dual AGN in UGC4211 to
  NanoHertz Gravitational Waves}},  \href{http://arxiv.org/abs/2306.17821}{{\tt
  arXiv:2306.17821}}.

\bibitem{Yang:2023aak}
J.~Yang, N.~Xie, and F.~P. Huang, {\it {Nano-Hertz stochastic gravitational
  wave background as hints of ultralight axion particles}},
  \href{http://arxiv.org/abs/2306.17113}{{\tt arXiv:2306.17113}}.

\bibitem{Guo:2023hyp}
S.-Y. Guo, M.~Khlopov, X.~Liu, L.~Wu, Y.~Wu, and B.~Zhu, {\it {Footprints of
  Axion-Like Particle in Pulsar Timing Array Data and JWST Observations}},
  \href{http://arxiv.org/abs/2306.17022}{{\tt arXiv:2306.17022}}.

\bibitem{Anchordoqui:2023tln}
L.~A. Anchordoqui, I.~Antoniadis, and D.~Lust, {\it {Fuzzy Dark Matter, the
  Dark Dimension, and the Pulsar Timing Array Signal}},
  \href{http://arxiv.org/abs/2307.01100}{{\tt arXiv:2307.01100}}.

\bibitem{Shen:2023pan}
Z.-Q. Shen, G.-W. Yuan, Y.-Y. Wang, and Y.-Z. Wang, {\it {Dark Matter Spike
  surrounding Supermassive Black Holes Binary and the nanohertz Stochastic
  Gravitational Wave Background}},  \href{http://arxiv.org/abs/2306.17143}{{\tt
  arXiv:2306.17143}}.

\bibitem{Li:2023yaj}
Y.~Li, C.~Zhang, Z.~Wang, M.~Cui, Y.-L.~S. Tsai, Q.~Yuan, and Y.-Z. Fan, {\it
  {Primordial magnetic field as a common solution of nanohertz gravitational
  waves and Hubble tension}},  \href{http://arxiv.org/abs/2306.17124}{{\tt
  arXiv:2306.17124}}.

\bibitem{Lambiase:2023pxd}
G.~Lambiase, L.~Mastrototaro, and L.~Visinelli, {\it {Astrophysical neutrino
  oscillations after pulsar timing array analyses}},
  \href{http://arxiv.org/abs/2306.16977}{{\tt arXiv:2306.16977}}.

\bibitem{Franciolini:2023pbf}
G.~Franciolini, A.~Iovino, Junior., V.~Vaskonen, and H.~Veermae, {\it {The
  recent gravitational wave observation by pulsar timing arrays and primordial
  black holes: the importance of non-gaussianities}},
  \href{http://arxiv.org/abs/2306.17149}{{\tt arXiv:2306.17149}}.

\bibitem{Liu:2023ymk}
L.~Liu, Z.-C. Chen, and Q.-G. Huang, {\it {Implications for the non-Gaussianity
  of curvature perturbation from pulsar timing arrays}},
  \href{http://arxiv.org/abs/2307.01102}{{\tt arXiv:2307.01102}}.

\bibitem{Peccei:1977hh}
R.~Peccei and H.~R. Quinn, {\it {CP Conservation in the Presence of
  Instantons}},  {\em Phys. Rev. Lett.} {\bf 38} (1977) 1440--1443.

\bibitem{Peccei:1977ur}
R.~Peccei and H.~R. Quinn, {\it {Constraints Imposed by CP Conservation in the
  Presence of Instantons}},  {\em Phys. Rev. D} {\bf 16} (1977) 1791--1797.

\bibitem{Abbott:1982af}
L.~Abbott and P.~Sikivie, {\it {A Cosmological Bound on the Invisible Axion}},
  {\em Phys. Lett. B} {\bf 120} (1983) 133--136.

\bibitem{Preskill:1982cy}
J.~Preskill, M.~B. Wise, and F.~Wilczek, {\it {Cosmology of the Invisible
  Axion}},  {\em Phys. Lett. B} {\bf 120} (1983) 127--132.

\bibitem{Dine:1982ah}
M.~Dine and W.~Fischler, {\it {The Not So Harmless Axion}},  {\em Phys. Lett.
  B} {\bf 120} (1983) 137--141.

\bibitem{Co:2017mop}
R.~T. Co, L.~J. Hall, and K.~Harigaya, {\it {QCD Axion Dark Matter with a Small
  Decay Constant}},  {\em Phys. Rev. Lett.} {\bf 120} (2018), no.~21 211602,
  [\href{http://arxiv.org/abs/1711.10486}{{\tt arXiv:1711.10486}}].

\bibitem{Freese:1990rb}
K.~Freese, J.~A. Frieman, and A.~V. Olinto, {\it {Natural inflation with pseudo
  - Nambu-Goldstone bosons}},  {\em Phys. Rev. Lett.} {\bf 65} (1990)
  3233--3236.

\bibitem{Dimopoulos:2005ac}
S.~Dimopoulos, S.~Kachru, J.~McGreevy, and J.~G. Wacker, {\it {N-flation}},
  {\em JCAP} {\bf 08} (2008) 003,
  [\href{http://arxiv.org/abs/hep-th/0507205}{{\tt hep-th/0507205}}].

\bibitem{Anber:2009ua}
M.~M. Anber and L.~Sorbo, {\it {Naturally inflating on steep potentials through
  electromagnetic dissipation}},  {\em Phys. Rev. D} {\bf 81} (2010) 043534,
  [\href{http://arxiv.org/abs/0908.4089}{{\tt arXiv:0908.4089}}].

\bibitem{Hook:2016mqo}
A.~Hook and G.~Marques-Tavares, {\it {Relaxation from particle production}},
  {\em JHEP} {\bf 12} (2016) 101, [\href{http://arxiv.org/abs/1607.01786}{{\tt
  arXiv:1607.01786}}].

\bibitem{Fonseca:2019ypl}
N.~Fonseca, E.~Morgante, R.~Sato, and G.~Servant, {\it {Axion fragmentation}},
  {\em JHEP} {\bf 04} (2020) 010, [\href{http://arxiv.org/abs/1911.08472}{{\tt
  arXiv:1911.08472}}].

\bibitem{Graham:2015cka}
P.~W. Graham, D.~E. Kaplan, and S.~Rajendran, {\it {Cosmological Relaxation of
  the Electroweak Scale}},  {\em Phys. Rev. Lett.} {\bf 115} (2015), no.~22
  221801, [\href{http://arxiv.org/abs/1504.07551}{{\tt arXiv:1504.07551}}].

\bibitem{Chadha-Day:2021uyt}
F.~Chadha-Day, {\it {Axion-like particle oscillations}},  {\em JCAP} {\bf 01}
  (2022), no.~01 013, [\href{http://arxiv.org/abs/2107.12813}{{\tt
  arXiv:2107.12813}}].

\bibitem{Cicoli:2012sz}
M.~Cicoli, M.~Goodsell, and A.~Ringwald, {\it {The type IIB string axiverse and
  its low-energy phenomenology}},  {\em JHEP} {\bf 10} (2012) 146,
  [\href{http://arxiv.org/abs/1206.0819}{{\tt arXiv:1206.0819}}].

\bibitem{Anber:2012du}
M.~M. Anber and L.~Sorbo, {\it {Non-Gaussianities and chiral gravitational
  waves in natural steep inflation}},  {\em Phys. Rev. D} {\bf 85} (2012)
  123537, [\href{http://arxiv.org/abs/1203.5849}{{\tt arXiv:1203.5849}}].

\bibitem{Co:2018lka}
R.~T. Co, A.~Pierce, Z.~Zhang, and Y.~Zhao, {\it {Dark Photon Dark Matter
  Produced by Axion Oscillations}},  {\em Phys. Rev. D} {\bf 99} (2019), no.~7
  075002, [\href{http://arxiv.org/abs/1810.07196}{{\tt arXiv:1810.07196}}].

\bibitem{Agrawal:2018vin}
P.~Agrawal, N.~Kitajima, M.~Reece, T.~Sekiguchi, and F.~Takahashi, {\it {Relic
  Abundance of Dark Photon Dark Matter}},  {\em Phys. Lett. B} {\bf 801} (2020)
  135136, [\href{http://arxiv.org/abs/1810.07188}{{\tt arXiv:1810.07188}}].

\bibitem{Dror:2018pdh}
J.~A. Dror, K.~Harigaya, and V.~Narayan, {\it {Parametric Resonance Production
  of Ultralight Vector Dark Matter}},  {\em Phys. Rev. D} {\bf 99} (2019),
  no.~3 035036, [\href{http://arxiv.org/abs/1810.07195}{{\tt
  arXiv:1810.07195}}].

\bibitem{Agrawal:2017eqm}
P.~Agrawal, G.~Marques-Tavares, and W.~Xue, {\it {Opening up the QCD axion
  window}},  {\em JHEP} {\bf 03} (2018) 049,
  [\href{http://arxiv.org/abs/1708.05008}{{\tt arXiv:1708.05008}}].

\bibitem{Machado:2018nqk}
C.~S. Machado, W.~Ratzinger, P.~Schwaller, and B.~A. Stefanek, {\it {Audible
  Axions}},  {\em JHEP} {\bf 01} (2019) 053,
  [\href{http://arxiv.org/abs/1811.01950}{{\tt arXiv:1811.01950}}].

\bibitem{Machado:2019xuc}
C.~S. Machado, W.~Ratzinger, P.~Schwaller, and B.~A. Stefanek, {\it
  {Gravitational wave probes of axion-like particles}},
  \href{http://arxiv.org/abs/1912.01007}{{\tt arXiv:1912.01007}}.

\bibitem{Geller:2021obo}
M.~Geller, S.~Lu, and Y.~Tsai, {\it {B modes from postinflationary
  gravitational waves sourced by axionic instabilities at cosmic
  reionization}},  {\em Phys. Rev. D} {\bf 104} (2021), no.~8 083517,
  [\href{http://arxiv.org/abs/2104.08284}{{\tt arXiv:2104.08284}}].

\bibitem{Figueroa:2023zhu}
D.~G. Figueroa, M.~Pieroni, A.~Ricciardone, and P.~Simakachorn, {\it
  {Cosmological Background Interpretation of Pulsar Timing Array Data}},
  \href{http://arxiv.org/abs/2307.02399}{{\tt arXiv:2307.02399}}.

\bibitem{Niu:2023bsr}
X.~Niu and M.~H. Rahat, {\it {NANOGrav signal from axion inflation}},
  \href{http://arxiv.org/abs/2307.01192}{{\tt arXiv:2307.01192}}.

\bibitem{Murai:2023gkv}
K.~Murai and W.~Yin, {\it {A Novel Probe of Supersymmetry in Light of Nanohertz
  Gravitational Waves}},  \href{http://arxiv.org/abs/2307.00628}{{\tt
  arXiv:2307.00628}}.

\bibitem{Madge:2023cak}
E.~Madge, E.~Morgante, C.~P. Ib\'a\~nez, N.~Ramberg, W.~Ratzinger, S.~Schenk,
  and P.~Schwaller, {\it {Primordial gravitational waves in the nano-Hertz
  regime and PTA data -- towards solving the GW inverse problem}},
  \href{http://arxiv.org/abs/2306.14856}{{\tt arXiv:2306.14856}}.

\bibitem{NANOGrav:2020bcs}
{\bf NANOGrav} Collaboration, Z.~Arzoumanian et~al., {\it {The NANOGrav 12.5 yr
  Data Set: Search for an Isotropic Stochastic Gravitational-wave Background}},
   {\em Astrophys. J. Lett.} {\bf 905} (2020), no.~2 L34,
  [\href{http://arxiv.org/abs/2009.04496}{{\tt arXiv:2009.04496}}].

\bibitem{Antoniadis:2022pcn}
J.~Antoniadis et~al., {\it {The International Pulsar Timing Array second data
  release: Search for an isotropic gravitational wave background}},  {\em Mon.
  Not. Roy. Astron. Soc.} {\bf 510} (2022), no.~4 4873--4887,
  [\href{http://arxiv.org/abs/2201.03980}{{\tt arXiv:2201.03980}}].

\bibitem{Adshead:2020htj}
P.~Adshead, G.~Holder, and P.~Ralegankar, {\it {BBN constraints on dark
  radiation isocurvature}},  {\em JCAP} {\bf 09} (2020) 016,
  [\href{http://arxiv.org/abs/2006.01165}{{\tt arXiv:2006.01165}}].

\bibitem{Planck:2018vyg}
{\bf Planck} Collaboration, N.~Aghanim et~al., {\it {Planck 2018 results. VI.
  Cosmological parameters}},  {\em Astron. Astrophys.} {\bf 641} (2020) A6,
  [\href{http://arxiv.org/abs/1807.06209}{{\tt arXiv:1807.06209}}]. [Erratum:
  Astron.Astrophys. 652, C4 (2021)].

\bibitem{Ratzinger:2020oct}
W.~Ratzinger, P.~Schwaller, and B.~A. Stefanek, {\it {Gravitational Waves from
  an Axion-Dark Photon System: A Lattice Study}},
  \href{http://arxiv.org/abs/2012.11584}{{\tt arXiv:2012.11584}}.

\bibitem{NANOGrav:2023tcn}
{\bf NANOGrav} Collaboration, G.~Agazie et~al., {\it {The NANOGrav 15 yr Data
  Set: Search for Anisotropy in the Gravitational-wave Background}},  {\em
  Astrophys. J. Lett.} {\bf 956} (2023), no.~1 L3,
  [\href{http://arxiv.org/abs/2306.16221}{{\tt arXiv:2306.16221}}].

\bibitem{Mingarelli:2017fbe}
C.~M.~F. Mingarelli, T.~J.~W. Lazio, A.~Sesana, J.~E. Greene, J.~A. Ellis,
  C.-P. Ma, S.~Croft, S.~Burke-Spolaor, and S.~R. Taylor, {\it {The Local
  Nanohertz Gravitational-Wave Landscape From Supermassive Black Hole
  Binaries}},  {\em Nature Astron.} {\bf 1} (2017), no.~12 886--892,
  [\href{http://arxiv.org/abs/1708.03491}{{\tt arXiv:1708.03491}}].

\bibitem{emcee}
D.~{Foreman-Mackey}, D.~W. {Hogg}, D.~{Lang}, and J.~{Goodman}, {\it emcee: The
  mcmc hammer},  {\em PASP} {\bf 125} (2013) 306--312,
  [\href{http://arxiv.org/abs/1202.3665}{{\tt arXiv:1202.3665}}].

\bibitem{Lewis:2019xzd}
A.~Lewis, {\it {GetDist: a Python package for analysing Monte Carlo samples}},
  \href{http://arxiv.org/abs/1910.13970}{{\tt arXiv:1910.13970}}.

\bibitem{Kitajima:2017peg}
N.~Kitajima, T.~Sekiguchi, and F.~Takahashi, {\it {Cosmological abundance of
  the QCD axion coupled to hidden photons}},  {\em Phys. Lett. B} {\bf 781}
  (2018) 684--687, [\href{http://arxiv.org/abs/1711.06590}{{\tt
  arXiv:1711.06590}}].

\bibitem{Eroncel:2022vjg}
C.~Er\"oncel, R.~Sato, G.~Servant, and P.~S\o{}rensen, {\it {ALP dark matter
  from kinetic fragmentation: opening up the parameter window}},  {\em JCAP}
  {\bf 10} (2022) 053, [\href{http://arxiv.org/abs/2206.14259}{{\tt
  arXiv:2206.14259}}].

\bibitem{Namba:2020kij}
R.~Namba and M.~Suzuki, {\it {Implications of Gravitational-wave Production
  from Dark Photon Resonance to Pulsar-timing Observations and Effective Number
  of Relativistic Species}},  {\em Phys. Rev. D} {\bf 102} (2020) 123527,
  [\href{http://arxiv.org/abs/2009.13909}{{\tt arXiv:2009.13909}}].

\bibitem{Heurtier:2021rko}
L.~Heurtier, F.~Huang, and T.~M.~P. Tait, {\it {Resurrecting low-mass axion
  dark matter via a dynamical QCD scale}},  {\em JHEP} {\bf 12} (2021) 216,
  [\href{http://arxiv.org/abs/2104.13390}{{\tt arXiv:2104.13390}}].

\bibitem{Bringmann:2023opz}
T.~Bringmann, P.~F. Depta, T.~Konstandin, K.~Schmidt-Hoberg, and C.~Tasillo,
  {\it {Does NANOGrav observe a dark sector phase transition?}},
  \href{http://arxiv.org/abs/2306.09411}{{\tt arXiv:2306.09411}}.

\bibitem{EPTA:2023sfo}
{\bf EPTA} Collaboration, J.~Antoniadis et~al., {\it {The second data release
  from the European Pulsar Timing Array - I. The dataset and timing analysis}},
   {\em Astron. Astrophys.} {\bf 678} (2023) A48,
  [\href{http://arxiv.org/abs/2306.16224}{{\tt arXiv:2306.16224}}].

\end{thebibliography}\endgroup
\bibliographystyle{JHEP}

\end{document}